\documentclass[amsmath,letterpaper,pre,twocolumn,longbibliography]{revtex4-1}
\usepackage{amssymb}
\usepackage{graphicx}
\usepackage{array}
\usepackage{hhline}
\usepackage{longtable}
\usepackage{mathtools}
\usepackage{color}

\newcommand{\wk}{\text{\tiny WK}}

\begin{document}

\title{Cellular signaling networks function 
  as generalized Wiener-Kolmogorov filters to suppress noise} \author{Michael Hinczewski
  and D. Thirumalai} \affiliation{Biophysics Program, Institute For
  Physical Science and Technology, University of Maryland, College
  Park, MD 20742}

\begin{abstract}
  Cellular signaling involves the transmission of environmental
  information through cascades of stochastic biochemical reactions, inevitably
  introducing noise that compromises signal fidelity.  Each stage of
  the cascade often takes the form of a kinase-phosphatase push-pull
  network, a basic unit of signaling pathways whose malfunction is
  linked with a host of cancers.  We show this ubiquitous enzymatic
  network motif effectively behaves as a Wiener-Kolmogorov (WK)
  optimal noise filter.  Using concepts from umbral calculus, we
  generalize the linear WK theory, originally introduced in the
  context of communication and control engineering, to take nonlinear
  signal transduction and discrete molecule populations into
  account. This allows us to derive rigorous constraints for efficient
  noise reduction in this biochemical system.  Our mathematical
  formalism yields bounds on filter performance in cases important to
  cellular function---like ultrasensitive response to stimuli.  We
  highlight features of the system relevant for optimizing filter
  efficiency, encoded in a single, measurable, dimensionless
  parameter.  Our theory, which describes noise control in a large
  class of signal transduction networks, is also useful both for the
  design of synthetic biochemical signaling pathways, and the
  manipulation of pathways through experimental probes like
  oscillatory input.
\end{abstract}

\maketitle
\def\s{\rule{0in}{0.28in}}

Extracting signals from time series corrupted by noise is a challenge
in a number of seemingly unrelated areas.  Minimizing the effects of
noise is a critical consideration in designing communication and
navigation systems, and analyzing data in diverse fields like medical
and astronomical imaging.  More recently, a number of studies have
focused on how biological circuits, comprised of chemical signaling
pathways mediated by genes, proteins, and RNA, cope with
noise~\cite{Altschuler2010}.  One of the key discoveries in the past
decade is that the naturally occurring systems that control all
aspects of cellular processes undergo substantial stochastic
fluctuations both in their expression levels and activities.  Noise
may even have a functional role~\cite{Cai08}, providing coordination
between multiple interacting chemical partners in typical circuits.
Because of the variety of ways noise influences cellular functions, it
is important to develop a practical and general theoretical framework
for describing how biological systems cope with and control the
inevitable presence of noise arising from stochastic fluctuations.  In
the context of communication theory, the optimal noise-reduction
filter, discovered independently by Wiener~\cite{Wiener49} and
Kolmogorov~\cite{Kolmogorov41} in the 1940's, inaugurated the modern
era of signal processing, providing the first general solution to the
problem of extracting useful information from corrupted signals.  We
show that this classic result of wartime mathematics, developed to
guide radar-assisted anti-aircraft guns, yields insights into the
efficiency limits of generic biochemical signaling networks.

\begin{figure*}[t]
\includegraphics[width=1.5\columnwidth]{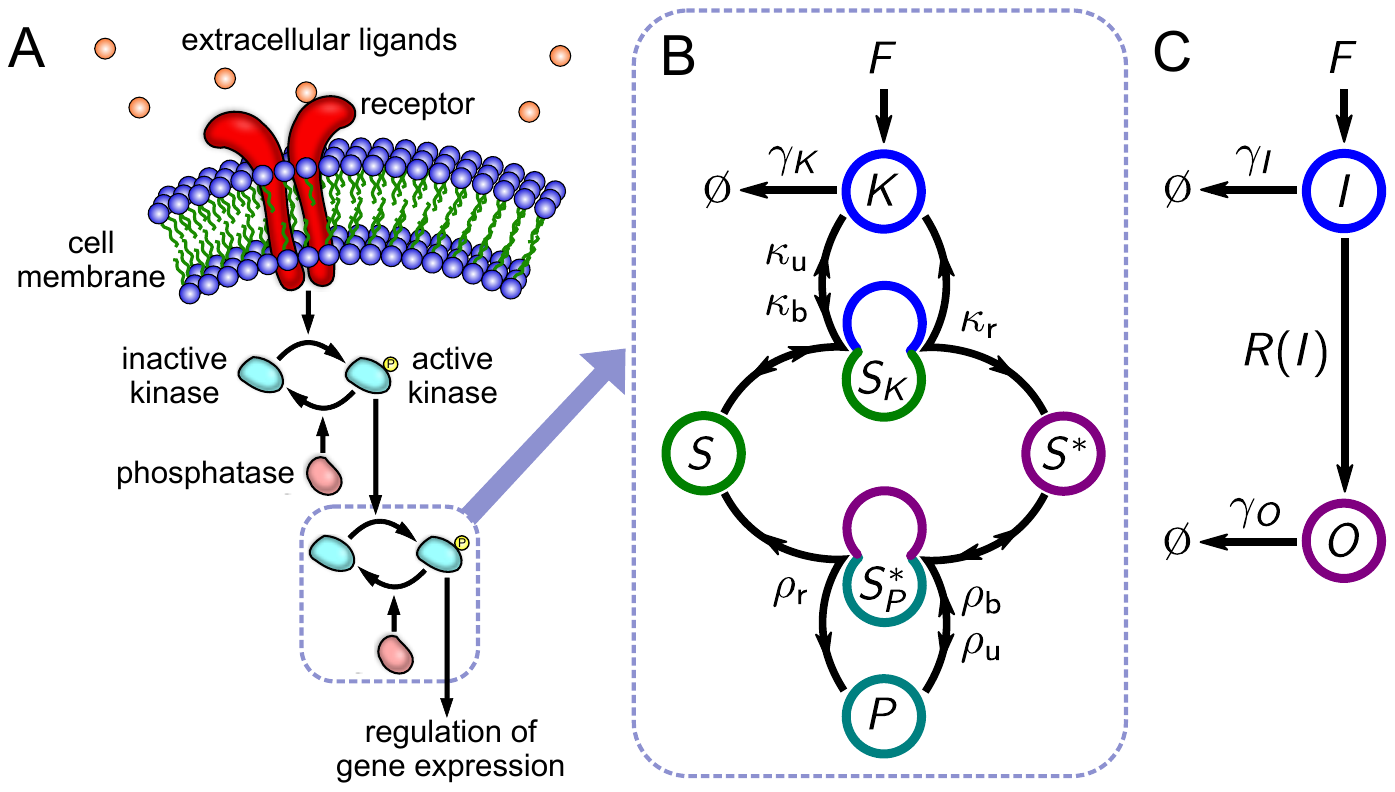}
\caption{Schematic of a signaling cascade.  A: A signaling pathway
  involving cascades of kinase phosphorylation, activated by a
  receptor embedded in the cell membrane which responds to
  extracellular ligands. B: A close-up of one enzymatic push-pull loop
  within the cascade.  Kinase ($K$) phosphorylates the substrate
  ($S$), converting it to active form ($S^\ast$), while phosphatase
  ($P$) reverts it to the original form through dephosphorylation.
  $S_K$ and $S^\ast_P$ represent the substrate in complex with the
  kinase and phosphatase respectively.  The rate parameters labeling
  the reaction arrows are described in the text.  The input $I =
  K+S_K$ and output $O = S^\ast+S^\ast_P$. C: A
  minimal signaling circuit, involving an input species $I$ and output
  species $O$, related by the production rate function
  $R(I)$.}\label{f1}
\end{figure*}

Dealing with noise in biological signal transduction is at first
glance even more daunting than in engineered systems.  In order to
survive, cells must process information about their external
environment~\cite{Mettetal08,Hersen08,Cheong11,Balazsi11Cell,Bowsher13},
which is transmitted and amplified from stimulated receptors on the
cell surface through elaborate pathways of post-translational covalent
modifications of proteins.  A typical example is phosphorylation by
protein kinases of target proteins, which then become activated to
modify targets further downstream.  Signaling occurs through cascades
involving multiple stages of such activation [Fig.~\ref{f1}A]. Since
each enzymatic reaction is stochastic, noise inevitably propagates
through the cascade, potentially corrupting the
signal~\cite{Thattai02,TanaseNicola06}.  Our work focuses on a basic
signaling circuit: a ``push-pull loop'' where a substrate is activated
by one enzyme (i.e. phosphorylation by a kinase) and deactivated by
another (i.e. dephosphorylation by a
phosphatase)~\cite{Stadtman77,Goldbeter81,Detwiler00,Heinrich2002}
[Fig.~\ref{f1}B].  Since cascades have a modular structure, formed
through many such loops in series and parallel, understanding the
stochastic properties at the single loop level is a prerequisite to
addressing the complex behavior of entire
pathways~\cite{Samoilov05,Levine07,GomezUribe07}.

The push-pull loop can act like an amplifier, taking the input
signal---the time-varying population of kinase---and approximately
reproducing it at larger amplitude through the output---the population
of active, phosphorylated substrate~\cite{Detwiler00}.  Depending on
the parameters, small changes in the input can be translated into
large (but noise-corrupted) output variations.  The amplification is
essential for sensitive response to external stimuli, but it must also
preserve signal content to be useful for downstream processes.  Thus,
the signaling circuit, despite operating in a noisy environment, needs
to maintain a high fidelity between output and amplified input.

From a design perspective, the natural question that arises is what
are the general constraints on filter efficiency?  Are there rigorous
bounds, which depend only on certain collective features of the
underlying biochemical network architecture?  Discovering such bounds
is important both to explain the metabolic costs of noise suppression
in biological systems~\cite{Lestas10}, and also for bioengineering
purposes.  In particular, for constructing synthetic signaling
networks, we would like to make the most efficient communication
pathway with a limited set of resources (free energy costs).

To answer these questions, using the enzymatic push-pull loop as an
example, we introduce a new mathematical framework, inspired by the
Wiener-Kolmogorov (WK) theory for optimal noise filtration.  The
original WK theory has restrictions that make it of limited utility in
the biological context---it assumes that the input and output are
continuous variables describing stationary stochastic processes.  More
critically, the filter is assumed to be linear.  Exploiting the power
of exact analytical techniques based on umbral calculus~\cite{Roman},
we overcome these limitations, thus generalizing the WK approach.
This crucial theoretical development enables us to provide a rigorous
solution to the filter optimization problem, taking into account
discrete populations and nonlinearity.  We can thus understand
constraints in biologically significant regimes of the push-pull loop
behavior, for example highly nonlinear, ``ultrasensitive''
amplification~\cite{Goldbeter81}.  Our theory predicts that optimality
can be realized by tuning phosphatase levels, which we verified
through simulations of a microscopic model of the loop reaction
network, including cases where the system is driven by an oscillatory
input~\cite{Mugler10}, which is relevant to recent experimental
probes~\cite{Mettetal08,Hersen08}.   The optimality is
  robust, with the filter operating at near-optimal levels even when
  the WK conditions are only approximately fulfilled, over a broad
  range of realistic parameter values.  Although illustrated using a
push-pull loop, the theory is applicable to a large class of signaling
networks, including more complex features such as negative feedback or
multi-site phosphorylation of substrates.

\section*{Results and Discussion}

{\bf Theoretical framework for a minimal signaling circuit.} To obtain
the central results, we start with an example which illustrates the
efficacy of the WK theory, and suggests a way to a more detailed,
realistic model of the enzymatic push-pull loop.  Consider a small
portion of a signaling pathway [Fig.~\ref{f1}C], involving two
chemical species: one with time-varying population $I(t)$ (the
``input''), and another one with population $O(t)$ (the ``output'')
whose production depends on $I(t)$.  These could be, for example, the
active, phosphorylated forms of two kinases within a signaling
cascade, with $O$ downstream of $I$.  The upstream part of the pathway
contributes an effective production rate $F$ for species $I$, which in
general can be time-dependent, though for now we will make $F$
constant.  The output $O$ is produced by a reaction, $I
\xrightarrow{R(I)} I + O$, with a rate $R(I(t))$ that depends on the
input.  The species are deactivated with respective rates $\gamma_I$
and $\gamma_O$, mimicking the role of the phosphatases.
The input $I(t)$ will vary over a characteristic time scale
$\gamma_I^{-1}$, fluctuating around the mean $\bar{I} = F/\gamma_I$.
The output deactivation rate sets the response time scale
$\gamma_O^{-1}$ over which $O(t)$ can react to changes in the input.
The dynamical equations, within a continuum, chemical Langevin (CL)
description~\cite{Gillespie00}, are given by:
\begin{equation}\label{e1}
\frac{dI}{dt} = F - \gamma_I I + n_I, \quad 
\frac{dO}{dt} = R(I) - \gamma_O O + n_O,
\end{equation}
where the additive noise contribution $n_\alpha(t) = \sqrt{2
  \gamma_\alpha \bar{\alpha}}\, \eta_\alpha(t)$, with $\alpha = I, O$
and $\bar{\alpha}$ denoting the mean of population $\alpha$.  The
function $\eta_\alpha(t)$ is Gaussian white noise with correlation
$\langle \eta_\alpha(t) \eta_{\alpha^\prime}(t^\prime)\rangle =
\delta_{\alpha\alpha^\prime} \delta(t-t^\prime)$.  The
$\langle\:\rangle$ brackets denote an average over the ensemble of all
possible noise realizations.

For small deviations $\delta \alpha(t) = \alpha(t) - \bar{\alpha}$
from the mean populations $\bar{\alpha}$, Eq.~\eqref{e1} can be solved
using a linear approximation, where we expand the rate function to
first order, $R(I(t)) \approx R_0 \bar{I} + R_1 \delta I(t)$, with
coefficients $R_0$, $R_1 >0$.  (We will return later to the issues of
nonlinearity and discrete populations.)  The result is:
\begin{align}
\delta I(t) &= \int_{-\infty}^t dt^\prime\, e^{-\gamma_I (t-t^\prime)} n_I(t^\prime),\label{e2}\\
\delta O(t) &= \int_{-\infty}^t dt^\prime\, \frac{R_1}{G} e^{-\gamma_O (t-t^\prime)} \left[G \delta I(t^\prime) + \frac{G}{R_1} n_O(t^\prime)\right],\nonumber
\end{align}
where in the second line we have introduced an arbitrary scaling
factor $G>0$ (to be defined below) inside the brackets, and divided
through by $G$ outside the brackets.  The solution for $\delta O(t)$
has the structure of a linear noise filter equation: $\tilde{s}(t) =
\int_{-\infty}^t dt^\prime H(t-t^\prime) c(t^\prime)$, with $c(t) =
s(t) + n(t)$.  In this analogy, we have a signal $s(t) \equiv G \delta
I(t)$ together with a noise term $n(t) \equiv GR^{-1} n_O(t)$ forming
a corrupted signal $c(t)$.  The output $\tilde{s}(t) \equiv \delta
O(t)$ is produced by convolving $c(t)$ with a linear filter kernel
$H(t) \equiv R_1 G^{-1} \exp(-\gamma_O t)$.  As a consequence of
causality, the integrals in Eq.~\eqref{e2} run over $t^\prime < t$, so
the filtered output $\tilde{s}(t)$ at any time $t$ depends only on
$c(t^\prime)$ from the past.

The utility of mapping the push-pull system onto a noise filter comes
from the application of WK theory, which is designed to solve a key
optimization problem: out of all possible causal, linear filters
$H(t)$, what is the optimal function $H_\text{\tiny WK}(t)$ that
minimizes the differences between the output $\tilde{s}(t)$ and input
$s(t)$ time series.  In our example, this means having $\delta O(t)$
reproduce as accurately as possible the scaled input signal $G \delta
I(t)$.  Specifically, we would like to minimize the relative
mean-squared error $E = \langle (\tilde{s} - s)^2\rangle/ \langle
s^2\rangle$.  For a particular $\delta I(t)$ and $\delta O(t)$, the
value of $E$ is smallest when $G = \langle (\delta O)^2\rangle
/\langle \delta O \delta I\rangle$, which we will use to define the
gain $G$.  In this case $E$ reduces to $E = 1 - \langle \delta O
\delta I\rangle^2/(\langle (\delta O)^2\rangle \langle(\delta
I)^2\rangle)$.  The great achievement of 
  Wiener~\cite{Wiener49} and Kolmogorov~\cite{Kolmogorov41} was to
show that $H_\wk$ satisfies the following Wiener-Hopf equation:
\begin{equation}\label{e3}
C_{cs}(t) = \int_{-\infty}^t dt^\prime\, H_\wk(t-t^\prime) C_{cc}(t^\prime),\quad t>0
\end{equation}
where $C_{xy}(t) \equiv \langle x(t^\prime)y(t^\prime+t)\rangle$ is
the correlation between points in time series $x$ and $y$, assumed to
depend only on the time difference $t$.  Given $C_{cs}$ and $C_{cc}$,
which are properties of the signal $s(t)$ and noise $n(t)$, it
is possible to solve Eq.~\eqref{e3} for $H_\wk$.  The corresponding
minimum value of the error $E$ is:
\begin{equation}\label{e4}
E_\wk = 1 - \frac{1}{C_{ss}(0)} \int_0^\infty dt\, H_\wk(t) C_{cs}(t).
\end{equation}

The solution of the Wiener-Hopf equation requires the following
correlation functions, which can be derived from Eq.~\eqref{e2}:
$C_{ss}(t) = C_{cs}(t) = G^2 \bar{I} \exp(-\gamma_I |t|)$, $C_{nn}(t)
= 2 G^2 \bar{I} \delta(t)/ (\gamma_I \Lambda)$, and $C_{cc}(t) =
C_{ss}(t) + C_{nn}(t)$, where the parameter $\Lambda \equiv R_1^2/(R_0
\gamma_I)$.  Plugging these into Eq.~\eqref{e3}, we can solve for the
optimal filter function by assuming a generic ansatz $H_\wk(t) =
\sum_{i=1}^{N} A_i \exp(-\lambda_i t)$, finding the unknown
coefficients $A_i$ and rate constants $\lambda_i$ by comparing the
left and right sides of the equation.  In our case, a single
exponential ($N=1$) is sufficient to exactly satisfy Eq.~\eqref{e3}
(see details in Appendix~\ref{a:wk}), and we get $H_\wk(t) =
\gamma_I(\sqrt{1+\Lambda}-1)\exp(-\gamma_I \sqrt{1+\Lambda}\, t)$.
The conditions for achieving WK optimality, $H(t) = H_\wk(t)$, are
then:
\begin{equation}\label{e5}
\gamma_O =  \gamma_I\sqrt{1+\Lambda}, \qquad G = \frac{R_1}{\gamma_I(\sqrt{1+\Lambda}-1)}.
\end{equation}
From Eq.~\eqref{e4} the minimum relative error is:
\begin{equation}\label{e6}
E_\wk = \frac{2}{1+\sqrt{1+\Lambda}}.
\end{equation}

The fidelity between output and input is described through a single
dimensionless optimality control parameter, $\Lambda$.  It can be
broken up into two multiplicative factors, reflecting two physical
contributions: $\Lambda = (R_0/\gamma_I) (R_1/R_0)^2$.  The first
term, $R_0/\gamma_I$, is a burst factor, measuring the mean number of
output molecules produced per input molecule during the active
lifetime of the input molecule.  The second term, $(R_1/R_0)^2$, is a
sensitivity factor, reflecting the local response of the production
function $R(I)$ near $\bar{I}$ (controlled by the slope $R_1 =
R^\prime(\bar{I})$) relative to the production rate per input molecule
$R_0 = R(\bar{I})/\bar{I}$.  Note that $(R_1/R_0)^2 >1$ only if $R(I)$
is globally nonlinear, since physical production functions satisfy
$R(I) \ge 0$ for all $I\ge0$.  If $R(I)$ is perfectly linear, $R(I) =
R_0 I$, then $R_1 = R_0$, and $(R_1/R_0)^2=1$.  Thus the limit of
efficient noise suppression, $\Lambda \gg 1$, where $E_\wk$ becomes
small, can be achieved by making the burst factor $R_0/\gamma_I\gg 1$
and/or enhancing the sensitivity $(R_1/R_0)^2 \gg 1$, at the cost of
introducing nonlinear effects (discussed in detail below).  For
optimality to be realized, we additionally need an appropriate
separation of scales [Eq.~\eqref{e5}] between the characteristic time
of variations in the input signal, $\gamma_I^{-1}$, and the response
time of the output, $\gamma_O^{-1}$.  The latter should be faster by a
factor of $\sqrt{1+\Lambda}$.  The scaling $E_\wk \sim \Lambda^{-1/2}$
for large $\Lambda$ is the same as the burst factor scaling of the
target population variance in biochemical negative feedback networks
intended to maintain homeostasis and suppress
fluctuations~\cite{Lestas10}.  The slow $\Lambda^{-1/2}$ decay in both
cases, compared to the more typical scaling of variance with
$\Lambda^{-1}$ (inversely proportional to the number of signaling
molecules produced) reflects the same underlying physical challenge:
the difficulty of suppressing or filtering noise in stochastic
reaction networks.

The error $E$ defined above is based on the
  instantaneous difference between the input $s(t)$ and output
  $\tilde{s}(t)$ time series.  One of the powerful features of the WK
  formalism is that it naturally extends error minimization to cases
  where the goal is extrapolating the future signal, where we seek to
  minimize the difference between $\tilde{s}(t)$ and $s(t+\alpha)$ for
  some $\alpha >0$~\cite{Bode50}.  Given the time delays inherent in
  many biological responses, particularly where feedback is involved,
  such predictive noise filtering has significant
  applications~\cite{Bialek2001}, which we will explore in subsequent
  work.  For now, we confine ourselves to the instantaneous error,
  which is sufficient to treat the kinase-phosphatase push-pull loop.

  We also note that there is no unique measure of signal
    fidelity.  Besides $E$, one can optimize the mutual information
    between the output and input species in the
    cascade~\cite{Levine07}.  For example, in the two-component
    cascade with nonlinear regulation, considered below, a spectral
    expansion of the master equation allows for efficient numerical
    optimization of the system parameters for particular forms of the
    rate function, maximizing the mutual
    information~\cite{Mugler09,Walczak09}.

{\bf Effects of nonlinearity and discrete populations.} For the
subclass of Gaussian-distributed signal $s(t)$ and noise $n(t)$ time
series (as is the case within the CL picture), the WK filter derived
above, based on the linearization of the CL, is optimal among all
possible linear or nonlinear filters~\cite{Bode50}.  If the system
fluctuates around a single stable state, and the copy numbers of the
species are large enough that their Poisson distributions converge to
Gaussians (mean populations $\gtrsim 10$), the signal and noise are
usually approximately Gaussian.  However, the rate function $R(I)$
will never be perfectly linear in practice, and thus one needs to
consider how nonlinearities in $R(I)$ will affect the minimal $E$. In
addition, the discrete nature of population changes, which becomes
important at lower copy numbers, has to be explicitly taken into
consideration.  Surprisingly, the WK result of Eq.~\eqref{e6} can be
generalized even to cases where the linear, continuum assumptions
underlying WK theory no longer hold.

Starting from the exact master equation, valid for discrete
populations and arbitrary $R(I)$, we have rigorously solved the
general optimization problem for the error $E$ between output and
input using the principles of umbral calculus~\cite{Roman}.  The
detailed proof is in Appendix~\ref{a:non}, but the main results are as follows.  Any
function $R(I)$ can be expanded in terms of a set of polynomials
$v_n(I)$ as $R(I) = \sum_{n=0}^\infty \sigma_n v_n(I)$.  The $v_n(I)$
are polynomials of degree $n$, given by
\begin{equation}\label{p1}
v_n(I) = \sum_{m=0}^n  (n-m)! \left(-\bar{I}\right)^m \binom{n}{m}\binom{I}{n-m},
\end{equation}
and the coefficients $\sigma_n$ are related to moments of $R(I)$,
$\sigma_n = \langle v_n(I) R(I)\rangle/(\bar{I}^n n!)$.  The average
is taken with respect to the Poisson distribution ${\cal P}(I) =
\bar{I}^I\exp(-\bar{I})/I!$.  The first two polynomials are $v_0 = 1$
and $v_1 = I-\bar{I}$, giving $\sigma_0 = \langle R(I) \rangle$ and
$\sigma_1 = \langle (I-\bar{I}) R(I) \rangle/\bar{I}$.  Remarkably,
the relative error $E$ has an exact analytical form in terms of the
$\sigma_n$,
\begin{equation}\label{p2}
E = 1 - \frac{\bar{I}\gamma_O^2 \sigma_1^2}{(\gamma_I+\gamma_O)^2}\left[\gamma_O \sigma_0 + \sum_{n=1}^\infty \sigma_n^2 \frac{\gamma_O n! \bar{I}^n}{\gamma_O+n\gamma_I} \right]^{-1}.
\end{equation}
This expression is bounded from below by
\begin{equation}\label{p3}
E \ge E_\text{opt} \equiv \frac{2}{1+\sqrt{1+\tilde \Lambda}},
\end{equation}
where $\tilde{\Lambda} = \bar{I} \sigma_1^2/(\sigma_0 \gamma_I)$.  The
equality is only reached when $\gamma_O = \gamma_I
\sqrt{1+\tilde{\Lambda}}$ and $R(I)$ has an optimal linear form,
$R_\text{opt}(I) = \sigma_0 + \sigma_1 (I-\bar{I})$, with all
$\sigma_{n}=0$ for $n \ge 2$.  In this optimal case, $\sigma_0 = R_0
\bar{I}$ and $\sigma_1 = R_1$, and hence $\tilde{\Lambda} = \Lambda$,
$E=E_\text{opt} = E_\wk$ from Eq.~\eqref{e6}.  

\begin{figure}[t]
\includegraphics[width=0.65\columnwidth]{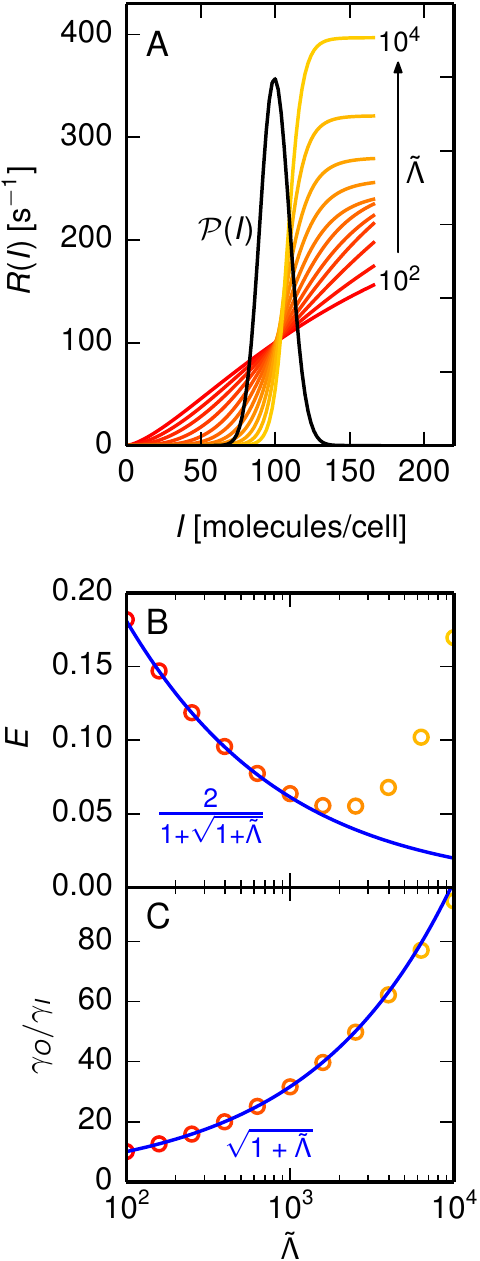}
\caption{Optimal noise reduction in the minimal signaling circuit
  (Fig.~\ref{f1}C).  A: Numerical optimization results for the Hill
  production function $R(I)$ that minimizes relative error $E$ between
  input and output, with each color corresponding to different values
  of the parameter $\tilde \Lambda=10^2-10^4$ (see text for other
  parameters).  The input probability distribution ${\cal P}(I)$ is
  superimposed in black (the height scale is arbitrary).  B: For each
  value of $\tilde \Lambda$ from panel A, circles show the minimal $E$.  The
  lower bound $E_\text{opt}$ [Eq.~\eqref{p3}] is drawn as a blue
  curve.  C: Analogous to panel B, but showing the ratio
    $\gamma_O/\gamma_I$ at which the minimum $E$ is achieved.  The
    blue curve shows the WK prediction for this ratio,
    $\gamma_O/\gamma_I = \sqrt{1+\tilde \Lambda}$.}\label{f2}
\end{figure}

Making $\tilde{\Lambda}$ large, for example by increasing $\sigma_1$,
is desirable for better signal transduction, but with a caveat.  We
can keep $E$ near $E_\text{opt}$ even for a globally nonlinear $R(I)$
so long as $R(I)$ remains approximately linear in the vicinity of the
mean $\bar{I}$, and the nonlinear corrections $\sigma_n$ for $n\ge 2$
are negligible.  Large $\sigma_1$ can be achieved through a highly
sigmoidal input-output response, known as ultrasensitivity, which is
biologically realizable in certain regimes of signaling
cascades~\cite{Goldbeter81}.  However, our theory predicts that as
$R(I)$ goes to the extreme limit of a step-like profile around
$\bar{I}$, $E$ should become significantly higher than $E_\text{opt}$,
and the benefits of ultrasensitivity vanish.   The reason
  for this is that letting $\sigma_1$ become arbitrarily large (making
  the step sharper) necessarily implies that $R(I)$ eventually
  deviates substantially from $R_\text{opt}(I)$.  We know that any
  physically sensible $R(I)$ satisfies the constraint $R(I) \ge 0$ for
  $I \ge 0$.  If $\sigma_1 \gg \sigma_0/\bar{I}$ and $\sigma_n \approx
  0$ for $n\ge 2$, the function $R(I)$ would be negative for $I
  \lesssim \bar{I} - \sigma_0/\sigma_1$, violating the physical
  constraint.  Hence the coefficients $\sigma_n$ for $n\ge 2$ must be
  non-negligible when $\sigma_1$ is sufficiently large, leading to $E
  > E_\text{opt}$.

  We can illustrate this result numerically for $R(I)$ that have the
  form of a Hill function, $R(I) = R_\text{s} (I/I_0)^{n_\text{H}} /
  (1+ (I/I_0)^{n_\text{H}})$, defined by the three parameters
  $R_\text{s}$, $I_0$, and ${n_\text{H}}$.  This represents a typical
  sigmoidal behavior in biochemical systems, with a small production
  rate for $I \ll I_0$ switching over to a saturation level
  $R_\text{s}$ for $I \gg I_0$.  We performed a numerical minimization
  of $E$ (evaluated using Eq.~\eqref{p2}) over the parameter space, at
  fixed $F$, $\gamma_I$, $\sigma_0$, and $\tilde \Lambda$.  Using
  Eq.~\eqref{p2} is numerically extremely efficient, since the
  coefficients $\sigma_n$ typically decay quite rapidly, allowing the
  infinite sum to converge after a small ($< 10$) number of terms.
  Fixing $\sigma_0$ and $\tilde \Lambda$ is equivalent to specifying
  the first two moments of $R(I)$, which in turn defines a curve in
  the three-dimensional parameter space of $R_\text{s}$, $I_0$, and
  ${n_\text{H}}$.  After numerically solving for this
    curve, the minimization procedure consists of searching along the
    curve (and varying the free system parameter $\gamma_O$) to find
    the parameter set that yields the smallest $E$.  Fig.~\ref{f2}A
  shows optimization results for $F=1$ s$^{-1}$, $\gamma_I = 0.01$
  s$^{-1}$, $\sigma_0 = 100$ s$^{-1}$, and varying $\tilde \Lambda$,
  with the optimal Hill function $R(I)$ (the one with smallest $E$) at
  each $\tilde \Lambda$ drawn in a different color.  The corresponding
  minimal values of $E$ are shown in Fig.~\ref{f2}B as circles in the
  same colors, with $E_\text{opt}$ using Eq.~\eqref{p3} drawn as a
  blue curve for comparison.  Larger values of $\tilde \Lambda$ have
  optimal $R(I)$ profiles that are increasingly step-like, with
  steeper slopes near $\bar{I}$.  For the range $\tilde \Lambda =10^2
  - 10^3$ the maximum slope ($\approx \sigma_1$) is still small enough
  that $R(I)$ remains approximately linear across the entire $I$ range
  where ${\cal P}(I)$ is non-negligible (the distribution is
  superimposed in Fig.~\ref{f2}A).  Hence minimal $E$ values are very
  close to $E_\text{opt}$, decreasing with $\tilde \Lambda$.
  The ratios $\gamma_O/\gamma_I$ at which these minimal
    $E$ values occur, shown in Fig.~\ref{f2}C, are nearly equal to the
    predicted value $\sqrt{1+\tilde \Lambda}$ (blue curve).  We can
  estimate that this near-optimality will persist up to $\sigma_1
  \approx \sigma_0/(3\sqrt{\bar{I}})$, since that is roughly the slope
  of an $R(I)$ that rises from zero near the left edge of ${\cal
    P}(I)$ (at $I \approx \bar{I} - 3\sqrt{\bar{I}}$) to a value of
  $\sigma_0$ at $I = \bar{I}$.  For $\sigma_1 \gtrsim
  \sigma_0/(3\sqrt{\bar{I}})$, or equivalently $\tilde \Lambda \gtrsim
  \sigma_0/(9\gamma_I) = 1.1 \times 10^3$, the nonlinearity of $R(I)$
  becomes appreciable around $\bar{I}$, distorting the output signal
  and leading to minimal $E$ noticeably larger than $E_\text{opt}$,
  and actually increasing with $\tilde \Lambda$.  Thus moving towards
  the ultrasensitive limit $\tilde \Lambda \to \infty$ is initially
  beneficial for noise filtering, but only up to a point: $R(I)$ does
  not have to be globally linear, but local linearity of $R(I)$ near
  $\bar{I}$, which can be satisfied readily, is best for accurate
  signal transduction.

\begin{figure*}[t]
\includegraphics[width=\textwidth]{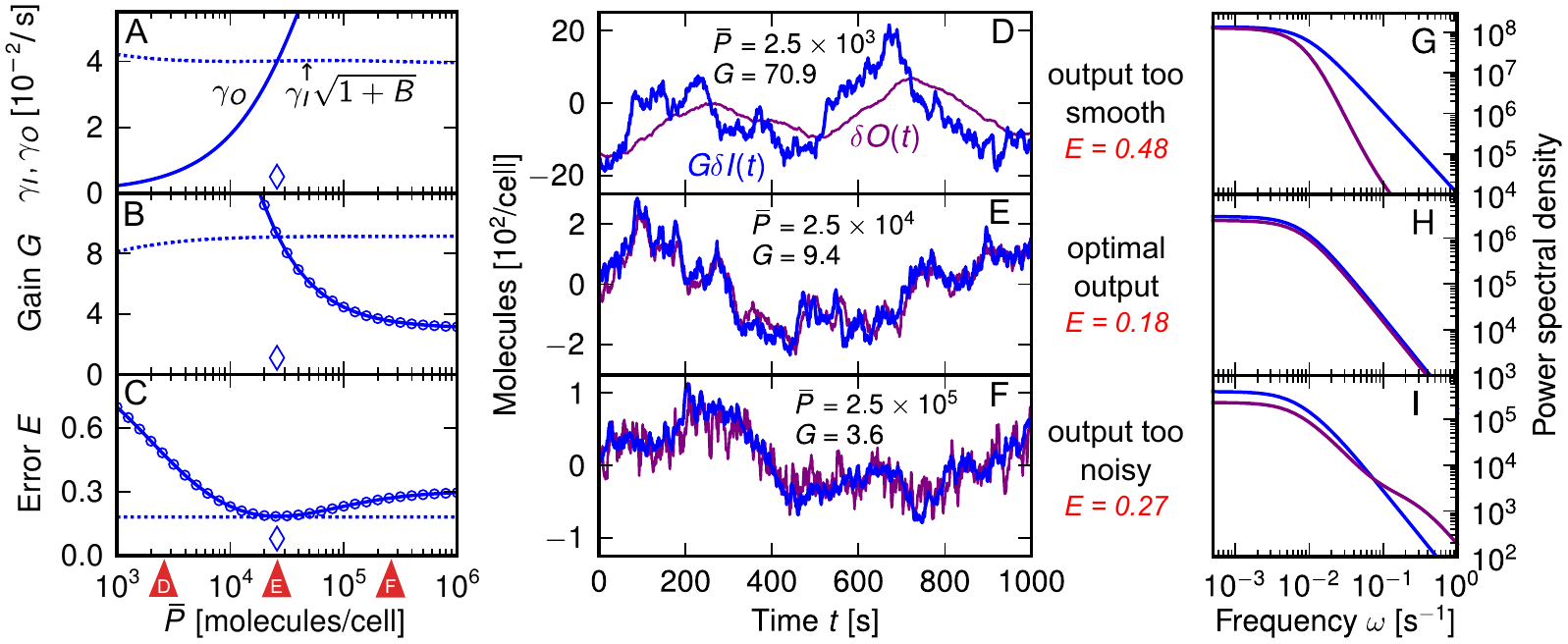}
\caption{A: $\gamma_O$ (solid curve) and $\gamma_I \sqrt{1+\Lambda}$
  (dashed curves), based on the mapping in Eq.~\eqref{e9}, for
  $\Lambda=100$.  The $\bar{P}$ value at the intersection of the solid
  and dashed curves, where the WK optimality conditions are fulfilled,
  is indicated by a diamond. B: Gain $G$ (solid line: CL theory;
  circles: KMC simulations) for the same $\Lambda$ value as in panel
  A, versus the WK optimal value for $G$ (dashed line) given by
  Eq.~\eqref{e5}. C: Same as panel B, but showing error $E$ versus the
  WK optimal prediction $E_\wk$ (dashed line) from
  Eq.~\eqref{e6}. D-F: Sample trajectories for the scaled input
  $G\delta I(t)$ (blue) and the output $\delta O(t)$ (purple) from KMC
  simulations of the push-pull loop, for $\Lambda=100$ ($\bar{S} = 8
  \times 10^4$) and three different values of $\bar{P}$.  These
  $\bar{P}$ values are marked by red triangles under panel C, and
  correspond to cases where, relative to the input, the output is too
  smooth (D), optimal (E), and too noisy (F).  G-I: Power spectral
  densities of the scaled input $G\delta I(t)$ (blue) and the output
  $\delta O(t)$ (purple) for the three cases shown in panels D-F.}\label{f3}
\end{figure*}

{\bf Enzymatic push-pull loop can act as an optimal WK filter.} The
system considered so far is the simplest realization of a signaling
circuit, in the sense that it involves only two species, related
through a single phenomenological production function, $R(I)$.  In
reality, an enzymatic push-pull loop involves
intermediates---complexes of the substrate with the kinase or
phosphatase---whose binding, unbinding, and catalytic reactions all
contribute to the stochastic nature of signal transmission.  Can the
WK theoretical framework be used to describe optimality in this
complicated context?  Let us consider a more microscopic model of the
loop reaction network [Fig.~\ref{f1}B].  The active kinase is either
free ($K$) or bound to substrate ($S_K$).  The input $I$ is defined as
the total active kinase population $I = K + S_K$.  Upstream modules
control kinase activation and deactivation, described by rates $F$ and
$\gamma_K$ respectively.  The kinase can phosphorylate the substrate,
converting it from inactive ($S$) to active ($S^\ast$) form.
Analogously, in the reverse direction, free phosphatases ($P$) form
complexes with the active substrate ($S^\ast_P$), which lead to
dephosphorylation, returning the substrate to inactive form.  The
output $O$ is the total active substrate population $O = S^\ast +
S^\ast_P$.  The reactions for substrate modification, with
corresponding rate constants, are:
\begin{equation}\label{e8}
\begin{split}
K + S
&\xrightleftharpoons[\kappa_\text{u}]{\kappa_\text{b}} S_K
\xrightarrow{\kappa_\text{r}} K + S^\ast\\
P + S^\ast
&\xrightleftharpoons[\rho_\text{u}]{\rho_\text{b}} S^\ast_P
\xrightarrow{\rho_\text{r}} P + S.
\end{split}
\end{equation}
We chose representative rate values based on a model of the MAP kinase
cascade~\cite{Schoeberl02} (all units are in s$^{-1}$):
$\kappa_\text{b} = \rho_\text{b} = 10^{-5}$, $\kappa_\text{u} = 0.02$,
$\rho_\text{u} = 0.5$, $\kappa_\text{r} = 3$, $\rho_\text{r} = 0.3$,
$F = 1$.  The rate $\gamma_K$ in the model controls the characteristic
time scale over which the input signal varies.  We let $\gamma_K =
0.01$ s$^{-1}$, which sets this scale to minutes.  Mean free substrate
and phosphatase populations (which together with the rates determine
all equilibrium population values) are in the ranges: $\bar{S} \sim
10^4 - 10^5$, $\bar{P} \sim 10^3 - 10^6$ molecules/cell.

We simulated the dynamics of this system numerically using kinetic
Monte Carlo (KMC)~\cite{Gillespie77}, with sample input and output
trajectories shown in Fig.~\ref{f3}D-F for $\bar{S} = 8 \times 10^4$
and three values of $\bar{P}$.  As the free phosphatase population is
varied, we see different degrees of signal fidelity, with the closest
match between $\delta O(t)$ and $G \delta I(t)$ for the intermediate
case in Fig.~\ref{f3}E.  Are we seeing behavior similar to an optimal
WK filter?  As detailed in Appendix~\ref{a:map}, we can
approximately map the phosphorylation cycle to a noise filter using
the same method as in our first example: starting from the full
dynamical equations in the linear CL approximation, we derive the
correlation functions required to solve the Wiener-Hopf relation,
Eq.~\eqref{e3}.  The effective parameters resulting from the mapping
are:
\begin{align}
\gamma_O &= \frac{\rho_\text{r} \rho_+}{\sqrt{\rho^2 - 2 \rho_\text{r} \rho_+}},\: R_1 = \frac{\kappa_\text{r} \kappa_+ \rho}{\kappa \sqrt{\rho^2 - 2 \rho_\text{r} \rho_+}},\:\gamma_I = \frac{\kappa_- \gamma_K}{\kappa},\nonumber \\
\Lambda &= \frac{\kappa_\text{r} \kappa_+ \kappa^2 \rho^2}{\gamma_K \kappa_- ( \rho^2 (\kappa^2 - \kappa_\text{r} \kappa_+) - \rho_\text{r} \rho_+ \kappa^2)},\label{e9}
\end{align}
where $\kappa_+ = \kappa_\text{b} \bar{S}$, $\kappa_- =
\kappa_\text{u}+ \kappa_\text{r}$, $\kappa = \kappa_+ + \kappa_-$,
$\rho_+ = \rho_\text{b} \bar{P}$, $\rho_- = \rho_\text{u}+
\rho_\text{r}$, and $\rho = \rho_+ + \rho_-$.  Eq.~\eqref{e9} is valid
in the regime $\bar{K} = F/\gamma_K \ll \bar{S}, \bar{P}$, with
corrections of order $\bar{K}/\bar{S}$ and $\bar{K}/\bar{P}$ shown in
Appendix~\ref{a:map}.  Such a mapping allows us to use WK results in
Eqs.~\eqref{e5} and \eqref{e6} to predict the conditions for
optimality and the minimal possible $E$.  Figs.~\ref{f3}A-B show the
left (solid lines) and right-hand (dashed lines) sides of both
conditions in Eq.~\eqref{e5} as a function of $\bar{P}$ for
$\Lambda=100$ ($\bar{S} = 8 \times 10^4$).  The $\bar{P}$ value at the
intersections, where the conditions are fulfilled, is marked by a
diamond.  Fig.~\ref{f3}C shows that exactly at this value $E$ achieves
a minimum, given by $E_\wk$ from Eq.~\eqref{e6} (dashed line).  The CL
approximation (solid curves) and KMC simulations (circles) are in
excellent agreement.  Thus, the phosphorylation cycle can indeed be
tuned to behave like an optimal WK noise filter, even for a realistic
signaling model.  In light of the mapping in Eq.~\eqref{e9}, we can
now understand the behavior of the trajectories in Fig.~\ref{f3}D-F,
which correspond to $\Lambda=100$.  In panel D, where $\bar{P} = 2.5
\times 10^3$, we have $\gamma_O \ll \gamma_I \sqrt{1+\Lambda}$
[Fig.~\ref{f3}A], and the output $\delta O(t)$ becomes excessively
smooth, since it cannot respond quickly enough to changes in the input
signal $G \delta I(t)$.  The corresponding power
  spectral density (PSD) of the output, shown in panel G, is smaller
  at high frequencies compared to the PSD of the input.  In panel F,
we have the opposite situation of $\gamma_O \gg \gamma_I
\sqrt{1+\Lambda}$ at $\bar{P} = 2.5 \times 10^5$.  The output response
is too rapid, generating additional noise that obscures the signal.
 In this case the output PSD (panel I) has an extra high
  frequency contribution relative to the input PSD. Panel E
represents the optimal intermediate $\bar{P} = 2.5 \times 10^4$, where
$\gamma_O = \gamma_I \sqrt{1+\Lambda}$ and the WK conditions are
fulfilled.  The input and output PSDs (panel H) are
  similar at all frequencies.

\begin{figure}[t]
\includegraphics[width=\columnwidth]{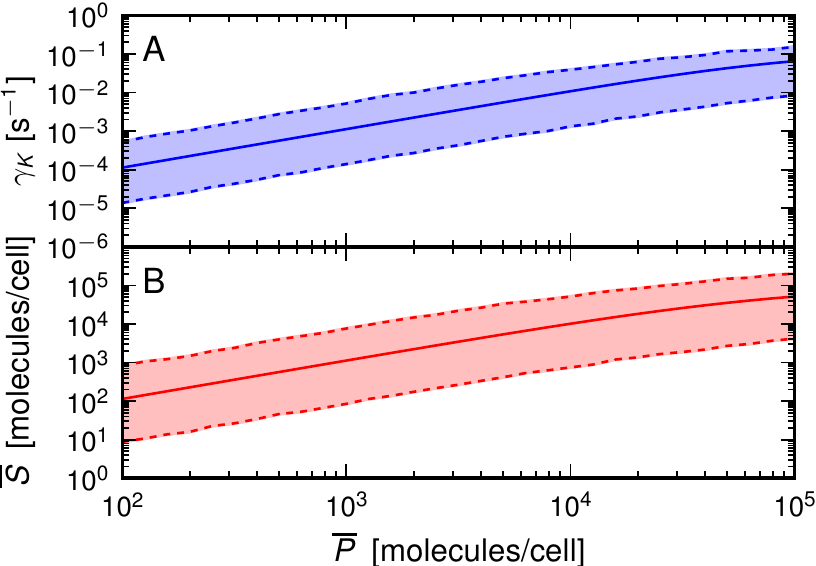}
\caption{Conditions for WK optimality as the enzymatic push-pull loop
  parameters are varied.  A: The solid blue curve shows the relation
  between mean phosphatase population $\bar{P}$ and the characteristic
  frequency scale $\gamma_K$ over which the active kinase input signal
  varies.  This is at WK optimality [Eq.~\eqref{e5}], using the
  mapping of Eq.~\eqref{e9} and the parameter values
  $\kappa_\text{b}$, $\rho_\text{b}$, $\kappa_\text{u}$,
  $\rho_\text{u}$, $\kappa_\text{r}$, $\rho_\text{r}$ listed in the
  text after Eq.~\eqref{e8}.  The shaded region between the dashed
  curves shows the 68\% confidence interval for achieving WK
  optimality, resulting from randomly perturbing all the parameter
  values so that they can be up to ten-fold smaller or larger.  B:
  Analogous to panel A, but showing the relation between $\bar{P}$ and
  substrate population $\bar{S}$ at WK optimality.}\label{f4}
\end{figure}

The minimum of $E$ in Fig.~\ref{f3}C is shallow, meaning that
near-optimal filtering persists even when the phosphatase population
is not precisely tuned to the WK condition.  For $\bar{P}$ values that
vary nearly five-fold between $\bar{P} = 1.3 - 6.3 \times 10^4$, the
error $E$ remains within 5\% of the minimum value $E_\wk$.  Another
aspect of the filter's robustness can be highlighted by perturbing the
enzymatic parameters $\kappa_\text{b}$, $\rho_\text{b}$,
$\kappa_\text{u}$, $\rho_\text{u}$, $\kappa_\text{r}$, and
$\rho_\text{r}$.  If we randomly vary all these parameters within a
range between 0.1 and 10 times the values listed above after
Eq.~\eqref{e8}, and calculate the resulting conditions for WK
optimality [Eq.~\eqref{e5}] for each new parameter set, we obtain the
results in Fig.~\ref{f4}.  For a given $\bar{P}$, the shaded intervals
in the figure correspond to the 68\% confidence intervals on the input
kinase frequency scale $\gamma_K$ and the mean substrate population
$\bar{S}$ at optimality.  Thus, for a broad range of biologically
relevant enzymatic parameters, we get a sense of how the populations
of $\bar{P}$ and $\bar{S}$ must complement each other, and an
associated time scale $\gamma_K^{-1}$ reflecting how quickly the input
signal can vary and still be accurately transduced.  From the trends
in Fig.~\ref{f4}, we see that to get the system to respond to more
rapidly varying signals, we need larger populations of $\bar{P}$ and
$\bar{S}$.  As a concrete example, for the hyperosmolar glycerol (HOG)
signaling pathway in yeast, discussed further in the next section,
kinase substrates have cell copy numbers of between $6 \times 10^1$
and $7 \times 10^3$, while the PTP and PTC phosphatases that have been
identified as targeting the pathway are present in cell copy numbers
between $1.5 \times 10^2$ and $2 \times 10^4$~\cite{Ghaemmaghami2003}.
Using these population scales as a rough guide for $\bar{S}$ and
$\bar{P}$ (ignoring complications like multiple phosphorylation
steps and sharing of phosphatases between different pathways) we see
from Fig.~\ref{f4} that the corresponding $\gamma_K \sim
10^{-4}-10^{-2}$ s$^{-1}$.  This range of optimal time scales is
consistent with the experimental observation that the HOG pathway can
faithfully transduce osmolyte signals at frequencies $\lesssim 5
\times 10^{-3}$ s$^{-1}$~\cite{Mettetal08}.

\begin{figure}[t]
\includegraphics[width=0.64\columnwidth]{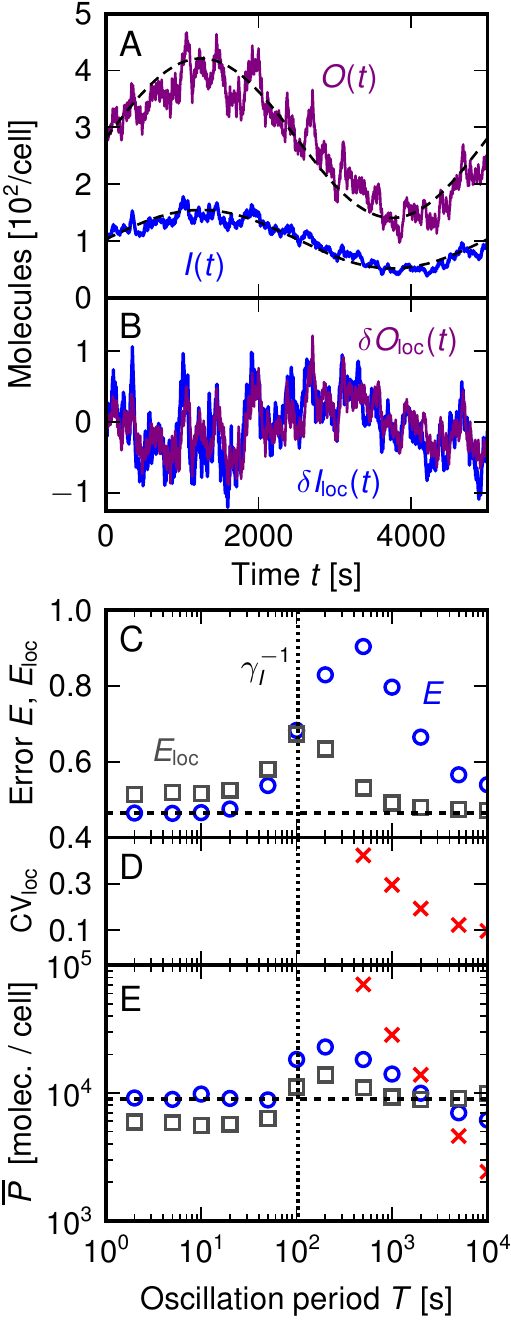}
\caption{A: sample KMC simulation trajectories for $I(t)$ (solid blue)
  and $O(t)$ (solid purple) in a $\Lambda=10$ system driven by an
  oscillatory upstream flux $F(t)$ (see text for parameters).  Dashed
  lines are local means $I_\text{loc}(t)$ and $O_\text{loc}(t)$.  B:
  for trajectories in A, the deviations from local means, $\delta
  I_\text{loc}(t)$ (blue) and $\delta O_\text{loc}(t)$ (purple).
  C-E: Results calculated from KMC for a system with
    $\Lambda=10$ and oscillatory $F(t)$ at varying driving periods
    $T$. $\gamma_I^{-1}$ is marked by a vertical dashed line.  C: The
    minimum errors $E$ (circles) and $E_\text{loc}$ (squares).
    $E_\wk$ is marked by a horizontal dashed line.  D: The minimum
    local coefficient of variation $\text{CV}_\text{loc} =
    \sqrt{\langle (\delta O_\text{loc}/ \bar{O}_\text{loc})^2
      \rangle}$.  E: The mean phosphatase population values $\bar{P}$
    at which the minima shown in panels C and D are achieved ($E$:
    circles, $E_\text{loc}$: squares, $\text{CV}_\text{loc}$:
    crosses).  The $\bar{P}$ value for WK optimality is marked by a
    horizontal dashed line.}\label{f5}
\end{figure}

{\bf Noise filtration in a push-pull loop driven by oscillatory
  input.}  Remarkably, since Eq.~\eqref{e9} is independent of $F$, the
system can serve as an optimal filter for a range of $F$ values, so
long as the condition $F/\gamma_K \ll \bar{S}, \bar{P}$ is satisfied.
This regime, involving saturated kinases and unsaturated phosphatases,
has been previously identified as a candidate for efficient signal
transmission by Gomez-Uribe {\em et al.}~\cite{GomezUribe07}.  To
check the filter operation with varying upstream flux, we used a
time-dependent $F(t)$, driving the system with oscillatory input.
This is motivated by microfluidic experimental
setups~\cite{Mettetal08,Hersen08}, where the HOG pathway of yeast was
probed by exposing the cells to periodic osmolyte pulses.  In the
experiments, the input signal is the extracellular osmolyte
concentration, and the output is the degree to which the activated
kinase Hog1 localizes in the nucleus, where it initiates a
transcriptional response to the osmolar shock.  Though the biochemical
network relating the output to input consists of a complex series of
enzymatic push-pull loops, the overall behavior was quantified through
response functions in terms of input signal
frequency~\cite{Mettetal08}, related to the Fourier transforms of the
input-output correlation functions.  Such correlation functions are
the basic ingredients in assessing filter optimality in the WK theory.
Here, we will focus only on a single push-pull loop, and use input at
varying frequencies to determine whether $E_\wk$ remains a meaningful
constraint on filter performance even for non-stationary signals.  In
Fig.~\ref{f5}A we show a sample $I(t)$ and $O(t)$ KMC trajectory at
optimality for $F(t) = \bar{F} (1+ A \sin (2\pi t/T))$, with $\bar{F}
=1$ s$^{-1}$, $A = 0.5$, and $T = 5000$ s.  The input has two
characteristic time scales, $T$ and $\gamma_I^{-1} = 10^2$ s.  For $T \gg
\gamma_I^{-1}$, we define relative error in terms of deviations from
local, time-dependent means: $E_\text{loc}$ defined using $\delta
I_\text{loc} = I(t) - \bar{I}_\text{loc}(t)$ and $\delta O_\text{loc}
= O(t) - \bar{O}_\text{loc}(t)$ [Fig.~\ref{f5}B], where
$\bar{I}_\text{loc}(t) = \bar{I} F(t)/\bar{F}$, $\bar{O}_\text{loc}(t)
= \bar{O} F(t)/\bar{F}$ are shown as dashed curves in Fig.~\ref{f5}A.
Fig.~\ref{f5}C shows KMC results for minimum $E$ and
  minimum $E_\text{loc}$ as a function of $T$ for a system tuned to
  optimality with $\Lambda=10$.  The values of $\bar{P}$ at which
  these minima are achieved are shown in Fig.~\ref{f5}E. At $T >
\gamma_I^{-1}$ we find $E_\text{loc} < E$, since both the input and
output have time to adjust to the slowly varying local means.  In
fact, the minimum $E_\text{loc}$ approaches $E_\wk$ for $T \gg
\gamma_I^{-1}$, as optimality is unaffected by the slow oscillation in
$F(t)$.  The $\bar{P}$ where the minimum $E_\text{loc}$
  occurs also approaches the value predicted by WK theory
  (Fig.~\ref{f5}E).  The filter transduces the signal with high
fidelity.  In the opposite limit of small $T < \gamma_I^{-1}$, the
rapidly varying $F(t)$ essentially averages out, since neither the
input nor the output have time to respond to the sharp changes in
$F(t)$.  Thus the system sees an effective constant flux $\bar{F}$.
Here $E$, the error estimate with respect to the global mean, is more
relevant than $E_\text{loc}$.  In this regime the minimum $E <
E_\text{loc}$, $E$ approaches $E_\wk$ for $T \ll \gamma_I^{-1}$, and
the $\bar{P}$ value where $E$ is minimized agrees with the WK
prediction.

The two regimes in system behavior, with a changeover at the time
scale $\gamma_I^{-1}$, reflect the fact that the enzymatic loop acts
an effective low-pass filter~\cite{Detwiler00,Hersen08}: it can
accurately transmit the low frequency component of $F(t)$, but
integrates over the high-frequency portion above a certain bandwidth.
The overall bandwidth of a cascade of push-pull loops has been
experimentally characterized for the yeast HOG pathway, yielding a
value of $\omega_b \approx 5 \times 10^{-3}$ s$^{-1}$~\cite{Hersen08}.
Using this as a rough estimate of the bandwidth scale $\gamma_I$ in
individual loops, we could expect to see a changeover between the two
regimes depending on whether the driving frequency is much slower or
faster than $\omega_b$.  Regardless of the magnitude of the driving
frequency, both $E$ and $E_\text{loc}$ always remain greater than
$E_\wk$, so the latter remains a bound on noise filter efficiency even
for dynamic input.

More generally, the low-pass filtering property of the
  enzymatic loop can be fine-tuned to optimize other signal
  transmission characteristics besides $E$ and $E_\text{loc}$.  These
  two errors are minimized when the output fluctuations ($\delta O$ or
  $\delta O_\text{loc}$) closely follow the scaled input fluctuations
  ($G \delta I$ or $G \delta I_\text{loc}$).  However one could
  imagine biological scenarios where the desired outcome was a
  smoothed output that mirrored the oscillatory driving signal.  In
  other words we could demand that $O(t)$, as shown for example in
  Fig.~\ref{f5}A (purple trajectory), deviates minimally from the
  oscillatory local mean $\bar{O}_\text{loc}(t)$ (superimposed dashed
  line).  In this case, the natural quantity to minimize would be a
  local coefficient of variation, $\text{CV}_\text{loc} =
  \sqrt{\langle (\delta O_\text{loc}/\bar{O}_\text{loc})^2 \rangle}$.
  From the oscillatory KMC simulations described above, we calculate
  $\text{CV}_\text{loc}$, and find that it can be made small in the
  slow oscillation regime $T \gg \gamma_I^{-1}$, as shown in
  Fig.~\ref{f5}D, which plots the minimum $\text{CV}_\text{loc}$ as a
  function of $T$ for $T \ge 500$ s.  From Fig.~\ref{f5}E, which shows
  the $\bar{P}$ values at which the minimum $\text{CV}_\text{loc}$
  occurs (crosses), we see that in the large $T$ limit this $\bar{P}$
  value is smaller than the WK prediction.  This makes sense, since as
  we know from the case of a constant driving function ($T \to
  \infty$), illustrated in Fig.~\ref{f3}D, keeping $\bar{P}$ below the
  WK optimum smooths the output.  For systems more complex than the
  enzymatic loop, smoothed output (homeostasis around a constant mean,
  or tracking of a driven, time-varying local mean) can be enhanced by
  introducing some negative feedback mechanism from the output back to
  the input~\cite{Lestas10}.  For such negative feedback systems it
  turns out there exists a mapping onto a different WK
  filter~\cite{HTpreprint}.

\section*{Conclusions}

We have demonstrated the usefulness of a generalized WK filter theory
as a way of characterizing signal fidelity in an enzymatic push-pull
loop.  This basic motif of biological signal transduction can
effectively realize an optimal WK noise filter.  Through a novel
analytical approach, we have generalized WK ideas beyond their
original linear context, thus providing fidelity bounds in strongly
nonlinear cases, including ultrasensitive production and oscillatory
input driving.  Even for a complex kinase-phosphatase reaction network
with multiple intermediates, the theory predicts the conditions for
accurate signal transduction, yielding a bound on the error in terms
of a single dimensionless optimality control parameter $\Lambda$.  The
results highlight how physics and engineering concepts can be use to understand 
how biology robustly tunes push-pull loops to
optimality by setting the copy numbers of phosphatase and substrate
molecules.  We can relate the wide range of cellular signaling protein
copy numbers observed experimentally to optimal time scales on which
the cell can accurately transduce the signal, and thus yield an
effective physiological response.  Since our approach is formulated in
terms of correlation functions of signal and noise, quantities readily
accessible from both theory and simulation, the current work can be
generalized to other complex signaling networks.  The ultimate goal is
to give insights into the design principles underlying the large,
intertwined biochemical pathways that determine how the cell can
process and respond to diverse sources of external stimuli.

\begin{acknowledgments}
  This work was supported by a grant from the National Science
  Foundation (CHE13-61946).
\end{acknowledgments}

\appendix

\section{Solving the Wiener-Hopf equation for the optimal filter}\label{a:wk}

Given the correlation functions,
\begin{equation}\label{s1}
\begin{split}
C_{ss}(t) &= C_{cs}(t) = G^2 \bar{I} e^{-\gamma_I |t|}, \quad C_{nn}(t) = \frac{2 G^2 \bar{I}}{\gamma_I \Lambda} \delta(t),\\
C_{cc}(t) &= C_{ss}(t) + C_{nn}(t),
\end{split}
\end{equation} 
we would like to find the optimal filter function $H_\wk(t)$ that
satisfies the Wiener-Hopf equation,
\begin{equation}\label{s2}
C_{cs}(t) = \int_{-\infty}^t dt^\prime\, H_\wk(t-t^\prime) C_{cc}(t^\prime),\quad t>0.
\end{equation}
Since $C_{cs}(t)$ and $C_{cc}(t)$ consist of exponential terms and
Dirac delta functions, a reasonable ansatz for $H_\wk(t)$ is a sum of $N$
exponentials, $H_\wk(t) = \sum_{i=1}^{N} A_i \exp(-\lambda_i t)$, with
parameters $A_i$, $\lambda_i$, $i=1,\ldots,N$.  Plugging this into
Eq.~\eqref{s2}, along with the correlation functions from
Eq.~\eqref{s1}, and carrying out the integral, we find
\begin{equation}\label{s3}
\begin{split}
G^2 \bar{I} e^{-\gamma_I t} &= \sum_{i=1}^N A_i \left[ \left(\frac{2 G^2\bar{I} \gamma_I}{\gamma_I^2 - \lambda_i^2} + \frac{2 G^2 \bar{I}}{\gamma_I \Lambda}\right)e^{-\lambda_i t}\right.\\
&\qquad \left. + \frac{G^2 \bar{I}}{\lambda_i - \gamma_I} e^{-\gamma_I t}\right], \quad t>0.
\end{split}
\end{equation}
Comparing the left-hand and right-hand sides of Eq.~\eqref{s3}, we see
that the coefficients of the linearly independent exponential terms on
both sides must match, giving $N+1$ equations: $N$ coefficients of
$\exp(-\lambda_i t)$, plus one for $\exp(-\gamma_I t)$.  Since there
are $2N$ unknown parameters in the ansatz, the only value of $N$ that
gives a closed set of equations is $N=1$.  With this choice of $N$,
the resulting two equations are
\begin{equation}\label{s4}
0 = A_1 \left(\frac{2 G^2\bar{I} \gamma_I}{\gamma_I^2 - \lambda_1^2} + \frac{2 G^2 \bar{I}}{\gamma_I \Lambda}\right), \qquad G^2 \bar{I} = \frac{A_1 G^2 \bar{I}}{\lambda_1 - \gamma_I}.
\end{equation}
The only physically sensible solution of Eq.~\eqref{s4} for $A_1$ and
$\lambda_1$ (where $|H_\wk(t)| \ne \infty$ as $t \to \infty$) is
\begin{equation}\label{s5}
A_1 = \gamma_I (\sqrt{1+\Lambda}-1), \quad \lambda_1 = \gamma_I \sqrt{1+\Lambda}.
\end{equation}
Thus the optimal filter is
\begin{equation}\label{s6}
H_\wk(t) = \gamma_I(\sqrt{1+\Lambda}-1)e^{-\gamma_I \sqrt{1+\Lambda}\, t}.
\end{equation}

\section{Optimal signal transduction for the nonlinear, discrete case}\label{a:non}

To obtain results for the general signal pathway model, where we
assume neither linearity of the production function $R(I)$ or a
continuum description, we start with an exact equation for the
stationary joint distribution ${\cal P}(I,O)$ of the input and output.
Using this, we will derive expressions for various moments of the
distribution which enter into the relative mean-squared error
\begin{equation}\label{s6b}
E = 1- \frac{\langle \delta O
\delta I\rangle^2}{\langle (\delta O)^2\rangle \langle (\delta I)^2
\rangle} = 1-\frac{(\langle O I \rangle - \langle O \rangle \langle I \rangle)^2}{(\langle O^2\rangle - \langle O \rangle^2)(\langle I^2 \rangle - \langle I \rangle^2)}.
\end{equation}
From the master equation, ${\cal P}(I,O)$ satisfies
\begin{equation}\label{s7}
\begin{split}
&\gamma_I \left[(I+1){\cal P}(I+1,O)-I{\cal P}(I,O)\right]\\
&\qquad +F \left[ {\cal P}(I-1,O)-{\cal P}(I,O)\right]\\
&\qquad + \gamma_O \left[(O+1){\cal P}(I,O+1)-O {\cal P}(I,O)\right]\\
&\qquad + R(I)\left[{\cal P}(I,O-1)-{\cal P}(I,O)\right]=0.
\end{split}
\end{equation}
Let us define a generating function $H_I(z) = \sum_{O=0}^\infty z^O
{\cal P}(I,O)$.  By multiplying Eq.~\eqref{s7} by $z^O$ and then
summing over $O$, we can derive the following equation for $H_I(z)$,
\begin{equation}\label{s8}
\begin{split}
&\gamma_I \left[(I+1)H_{I+1}(z)-I H_I(z)\right] +F \left[ H_{I-1}(z)-H_I(z)\right]\\
&\qquad+ \gamma_O (1-z)H^\prime_I(z) + R(I)(z-1)H_I(z)=0.
\end{split}
\end{equation}
Plugging in $z=1$, Eq.~\eqref{s8} can be solved for $H_I(1) = {\cal
  P}(I)$, the marginal probability distribution of the input.  The
result is ${\cal P}(I) = (F/\gamma_I)^I \exp(-F/\gamma_I)/I!$, the
Poisson distribution.  This implies that the first and second input
moments are given by
\begin{equation}\label{s8b}
\langle I \rangle = \frac{F}{\gamma_I} \equiv \bar{I}, \qquad \langle I^2\rangle = \frac{F^2}{\gamma_I^2} + \frac{F}{\gamma_I} = \bar{I}^2 + \bar{I}.
\end{equation}
Moments involving the output $O$ can be obtained by
manipulation of Eq.~\eqref{s8}.  Taking its first derivative with
respect to $z$, and then setting $z=1$, we find
\begin{equation}\label{s9}
\begin{split}
&\gamma_I \left[(I+1)H^\prime_{I+1}(1)-I H_I^\prime(1)\right] +F \left[ H^\prime_{I-1}(1)-H_I^\prime(1)\right]\\
&\qquad - \gamma_O H^\prime_I(1) + R(I)H_I(1)=0.
\end{split}
\end{equation}
Similarly, taking the second derivative of Eq.~\eqref{s8} with respect
to $z$, and setting $z=1$, gives
\begin{equation}\label{s10}
\begin{split}
&\gamma_I \left[(I+1)H^{\prime\prime}_{I+1}(1)-I H_I^{\prime\prime}(1)\right] +F \left[ H^{\prime\prime}_{I-1}(1)-H_I^{\prime\prime}(1)\right]\\
&\qquad- 2\gamma_O H^{\prime\prime}_I(1) + 2 R(I)H^\prime_I(1)=0.
\end{split}
\end{equation}
From the definition of the generating function, $H^\prime_I(1) =
\sum_{O=0}^\infty O {\cal P}(I,O)$ and $H^{\prime\prime}_I(1) =
\sum_{O=0}^\infty O(O-1) {\cal P}(I,O)$.  Summing Eqs.~\eqref{s9} and
\eqref{s10} over all $I$ yields the following moment relations,
\begin{equation}\label{s11}
\begin{split}
&\sum_{I=0}^\infty H^\prime_I(1) = \gamma_O^{-1}\sum_{I=0}^\infty R(I) H_I(1) \\
 &\qquad \Rightarrow \: \langle O \rangle = \gamma_O^{-1} \langle R(I) \rangle,\\
&\sum_{I=0}^\infty H^{\prime\prime}_I(1) = \gamma_O^{-1}\sum_{I=0}^\infty R(I) H^\prime_I(1) \\
&\qquad \Rightarrow \: \langle O^2 \rangle - \langle O \rangle= \gamma_O^{-1} \langle O R(I) \rangle.
\end{split}
\end{equation}
Evaluating $\langle O \rangle$ involves finding the mean of $R(I)$
over the known input distribution $H_I(1) = {\cal P}(I)$.  However,
finding $\langle O^2 \rangle$ involves the unknown distribution
$H^\prime_I(1)$.  Moreover, the last remaining moment in
Eq.~\eqref{s6b} for the mean-squared error, $\langle O I \rangle$, can
also be expressed in terms of this distribution, $\langle O I \rangle
= \sum_{I=0}^\infty I H^\prime_I(1)$.  Thus it is crucial to have
additional information about $H^\prime_I(1)$.

We know that $H^\prime_I(1)$ satisfies Eq.~\eqref{s9}, and let us
assume an ansatz for $H^\prime_I(1)$ of the form $H^\prime_I(1) =
\gamma_O^{-1} H_I(1) G(I)$ for some function $G(I)$.  Plugging this
into Eq.~\eqref{s9}, and using the fact that $H_I(1)$ is the Poisson
distribution, we find
\begin{equation}\label{s12}
H_I(1) \left[ ({\cal S}-1) G(I) + R(I) \right]=0,
\end{equation}
where ${\cal S}$ is an operator acting on $G(I)$, defined as
\begin{equation}\label{s13}
{\cal S} = \gamma_O^{-1}(\gamma_I I \Delta_{-1} + F\Delta_1).
\end{equation}
Here $\Delta_h$ is the finite difference operator, which acts on a
function $f(I)$ as $\Delta_h f(I) \equiv f(I+h)-f(I)$.  Thus the
function $G(I)$ which solves Eq.~\eqref{s12} is $G(I) = (1-{\cal
  S})^{-1} R(I) \equiv {\cal L} R(I)$, where the operator ${\cal L} =
\sum_{n=0}^\infty {\cal S}^n$.  Thus $H^\prime_I(1) = \gamma_O^{-1}
H_I(1) {\cal L} R(I)$, and
\begin{equation}\label{s13b}
\langle O I \rangle = \gamma_O^{-1}\langle I {\cal L} R(I)\rangle, \qquad \langle O R(I)\rangle = \gamma_O^{-1} \langle R(I) {\cal L} R(I)\rangle.
\end{equation}
Note that the terms on the right-hand sides inside the $\langle\,
\rangle$ brackets are solely functions of $I$, and hence the averages
depend on ${\cal P}(I)$.  Plugging Eqs.~\eqref{s8b},\eqref{s11}, and
\eqref{s13b} into Eq.~\eqref{s6b} gives an expression for the relative
error,
\begin{equation}\label{s13c}
\begin{split}
E &= 1 -\frac{\bar{I}^{-1} \langle (I {\cal L}-\bar{I}) R(I) \rangle^2}{\gamma_O\langle R(I) \rangle + {\cal M}[R(I)]}, \\
{\cal M}[R(I)] &\equiv \langle R(I) {\cal L} R(I) \rangle - \langle R(I) \rangle^2.
\end{split}
\end{equation}

To make further progress on the evaluation of $E$, it would be helpful
to express $R(I)$ in terms of eigenfunctions of ${\cal S}$ (which
would also be eigenfunctions of ${\cal L}$).  To do this, we employ a
set of techniques known as umbral calculus~\cite{Roman}, which starts
with the observation that the function $R(I)$ can be expanded in a
Newton series (the finite difference analogue of the Taylor series),
\begin{equation}\label{s14}
R(I) = \sum_{m=0}^\infty \rho_m (I)_m, \qquad \rho_m \equiv \frac{1}{m!} \left.\Delta^m_1 R(I)\right|_{I=0},
\end{equation}
where $(I)_m \equiv I(I-1)\cdots(I-m+1) = m! \binom{I}{m}$ is the
$m$th falling factorial of $I$ (with $(I)_0 \equiv 1$).
 The Newton series expansion exists assuming $R(I)$
  fulfills certain analyticity and growth conditions~\cite{Gelfond}, which are
  satisfied for all physically realistic production functions.
Finite difference operators acting on $(I)_m$ result in linear
combinations of falling factorials.  In particular, $\Delta_1 (I)_m =
m(I)_{m-1}$ and $I \Delta_{-1} (I)_m = -m (I)_m$.  Thus the operator
${\cal S}$ acting on $(I)_m$ gives
\begin{equation}\label{s15}
{\cal S} (I)_m = -\frac{m \gamma_I}{\gamma_O} \left[(I)_m - \bar{I} (I)_{m-1} \right].
\end{equation}
If we consider functions like $R(I)$ as vectors in the basis of
falling factorials $\{(I)_m,\,m=0,1,\ldots\}$, with components
$\rho_m$, then from Eq.~\eqref{s15} the operator ${\cal S}$ is a simple
bidiagonal matrix in this basis, with elements
\begin{equation}\label{s16}
{\cal S}_{m^\prime,m} = -\frac{m \gamma_I}{\gamma_O} \delta_{m^\prime,m} + \frac{m \gamma_I \bar{I}}{\gamma_O} \delta_{m^\prime,m-1}.
\end{equation}
The eigenvalues of $\lambda_n$ of ${\cal S}$, labeled by $n = 0,1,\ldots$ in
decreasing order, are just the diagonal matrix components, $\lambda_n
= -n \gamma_I/\gamma_O$.  The corresponding eigenfunctions are
\begin{equation}\label{s17}
v_n(I) = \sum_{m=0}^n  \binom{n}{m} \left(-\bar{I}\right)^m  (I)_{n-m}.
\end{equation}
The $n$th eigenfunction $v_n(I)$ is a polynomial in $I$ of degree $n$,
with the first few given by
\begin{equation}\label{s18}
\begin{split}
v_0(I) &= 1, \quad v_1(I) = I - \bar{I}, \quad v_2(I) = (I-\bar{I})^2 - I,\\
 v_3(I) &= (I-\bar{I})^3-3 I (I-\bar{I})+2 I.
\end{split}
\end{equation}
The eigenfunctions $v_n(I)$ are mathematically related to expansions of the master equation through alternative approaches, for example the spectral
  method of Refs.~\cite{Mugler09,Walczak09}.  In fact, $v_n(I) = n!
  \langle n | I \rangle$, where $\langle n|I \rangle$ is the mixed
  product defined in Eq.~A8 of Ref.~\cite{Mugler09} (with $\bar{I}$
  substituted for the rate parameter $g$).

Since Eq.~\eqref{s17} can be inverted to express $(I)_m$ in terms of
the eigenfunctions,
\begin{equation}\label{s19}
(I)_m = \sum_{n=0}^m \binom{m}{n} \bar{I}^{m-n} v_n(I),
\end{equation}
we can write $R(I)$ as in terms of the eigenfunctions by plugging
Eq.~\eqref{s19} into Eq.~\eqref{s14},
\begin{equation}\label{s20}
R(I) = \sum_{n=0}^\infty \sigma_n v_n(I), \qquad \sigma_n \equiv \sum_{m=0}^\infty \binom{m}{n}\rho_m \bar{I}^{m-n}, 
\end{equation}
where we have used the property that $\binom{m}{n} = 0$ for $n >m$.
The operator ${\cal L} = \sum_{k=0}^\infty {\cal S}^k$ acting on
$R(I)$ is then
\begin{equation}\label{s21}
\begin{split}
{\cal L} R(I) &= \sum_{n=0}^\infty \sigma_n \sum_{k=0}^\infty \left(-\frac{n \gamma_I}{\gamma_O}\right)^k v_n(I) \\
&= \sum_{n=0}^\infty \sigma_n \frac{\gamma_O}{\gamma_O + n \gamma_I} v_n(I).
\end{split}
\end{equation}
Since the quantities in Eq.~\eqref{s13c} for $E$ involve averages with
respect to ${\cal P}(I)$, it is useful to derive the first and second
moments of the eigenfunctions.  From the fact that the falling
factorials have very simple averages in the Poisson distribution,
$\langle (I)_m \rangle = \bar{I}^m$, we find using Eq.~\eqref{s17}
that $\langle v_n(I) \rangle = \delta_{n,0}$.  This implies that
$\langle R(I) \rangle = \langle {\cal L} R(I) \rangle = \sigma_0$.  To
find $\langle v_{n^\prime}(I) v_n(I) \rangle$, we start from the
Chu-Vandermonde identity~\cite{Roman}, the umbral analogue of the
binomial theorem,
\begin{equation}\label{s22}
(x+y)_m = \sum_{k=0}^m \binom{m}{k} (x)_{m-k}(y)_k.
\end{equation}
For $x=I-m^\prime$ and $y=m^\prime$ this gives
\begin{equation}\label{s22}
\begin{split}
(I)_m &= \sum_{k=0}^n \binom{m}{k} (I-m^\prime)_{m-k}(m^\prime)_k \\
&= \sum_{k=0}^n k! \binom{m}{k} \binom{m^\prime}{k} (I-m^\prime)_{m-k},
\end{split}
\end{equation}
where we have used the fact that $(m)_k = k!
\binom{m}{k}$. Multiplying both sides by $(I)_{m^\prime}$, we find
\begin{equation}\label{s23}
\begin{split}
(I)_{m^\prime}(I)_m  &= \sum_{k=0}^n k! \binom{m}{k} \binom{m^\prime}{k} (I)_{m^\prime}(I-m^\prime)_{m-k}\\
& = \sum_{k=0}^n k! \binom{m}{k} \binom{m^\prime}{k} (I)_{m+m^\prime-k}.
\end{split}
\end{equation}
The second equality is based on the relation $(I)_{i+j} = (I)_i
(I-i)_j$, which follows from the definition of the falling factorial.
Taking the average of both sides of Eq.~\eqref{s23} yields
\begin{equation}\label{s24}
\langle (I)_{m^\prime}(I)_m \rangle  = \sum_{k=0}^n k! \binom{m}{k} \binom{m^\prime}{k} \bar{I}^{m+m^\prime-k}.
\end{equation}
An alternative expression for $\langle (I)_{m^\prime}(I)_m \rangle$
can be derived by substituting the eigenfunction expansion of Eq.~\eqref{s19} for both $(I)_{m^\prime}$ and $(I)_m$,
\begin{equation}\label{s25}
\begin{split}
&\langle (I)_{m^\prime}(I)_m \rangle  = \\
&\quad\sum_{n^\prime=0}^{m^\prime}\sum_{n=0}^m \binom{m^\prime}{n^\prime}\binom{m}{n} \bar{I}^{m+m^\prime-n-n^\prime} \langle v_{n^\prime}(I) v_n(I)\rangle.
\end{split}
\end{equation}
Comparing the right-hand sides of Eqs.~\eqref{s24} and \eqref{s25} we
see that $\langle v_{n^\prime}(I) v_n(I)\rangle = n! \bar{I}^n
\delta_{n^\prime,n}$.  Together with Eqs.~\eqref{s20} and \eqref{s21}
this allows us to calculate
\begin{equation}\label{s26}
\begin{split}
{\cal M}[R(I)] &= \langle R(I) {\cal L} R(I) \rangle - \langle R(I) \rangle^2\\
 &= \sum_{n^\prime=0}^\infty \sum_{n=0}^\infty \sigma_{n^\prime}\sigma_n \frac{\gamma_O}{\gamma_O+n\gamma_I} \langle v_{n^\prime}(I) v_n(I)\rangle - \sigma_0^2\\
&=  \sum_{n=1}^\infty \sigma_n^2 \frac{\gamma_O n! \bar{I}^n}{\gamma_O+n\gamma_I}.
\end{split}
\end{equation}
Using the fact that $I = \bar{I} v_0(I) + v_1(I)$, we can similarly
evaluate
\begin{equation}\label{s27}
\begin{split}
&\langle (I {\cal L}-\bar{I}) R(I) \rangle\\
& = \sum_{n=0}^\infty \sigma_n \left[\bar{I}\langle v_0(I) v_n(I) \rangle + \frac{\gamma_O}{\gamma_O+\gamma_I}\langle v_1(I) v_n(I)\rangle\right] -\bar{I} \sigma_0\\
& = \frac{\gamma_O \bar{I} \sigma_1}{\gamma_O+\gamma_I}. 
\end{split}
\end{equation}
Plugging Eqs.~\eqref{s26} and \eqref{s27} into Eq.~\eqref{s13c}, we
obtain our final expression for the relative error,
\begin{equation}\label{s28}
E = 1 - \frac{\bar{I}\gamma_O^2 \sigma_1^2}{(\gamma_I+\gamma_O)^2}\left[\gamma_O \sigma_0 + \sum_{n=1}^\infty \sigma_n^2 \frac{\gamma_O n! \bar{I}^n}{\gamma_O+n\gamma_I} \right]^{-1}.
\end{equation}
This expression can be readily calculated numerically for any given
$R(I)$, as was done in the main text for the family of Hill function
production rates.  To facilitate evaluation, we express the
coefficients $\sigma_n$ as moments with respect to the Poisson
distribution ${\cal P}(I)$ in the following manner, using the
expansion of Eq.~\eqref{s20},
\begin{equation}\label{s28b}
\begin{split}
&\langle v_n(I) R(I) \rangle = \sum_{n^\prime=0}^\infty \sigma_{n^\prime} \langle v_{n^\prime}(I) v_{n}(I) \rangle = \sigma_n n! \bar{I}^n\\
&\qquad \Rightarrow \quad \sigma_n = \frac{\langle v_n(I) R(I) \rangle}{n! \bar{I}^n}.
\end{split}
\end{equation}
From the definition of $v_n(I)$ in Eq.~\eqref{s17}, the coefficients
$\sigma_n$ can be written
\begin{equation}\label{s28c}
\sigma_n = \sum_{m=0}^n \frac{(-1)^{n-m}\bar{I}^{-m}}{(n-m)!} \left\langle \binom{I}{m} R(I) \right\rangle.
\end{equation}
Using Eq.~\eqref{s28c} the $\sigma_n$ can be numerically calculated
for any $R(I)$.  The sum in Eq.~\eqref{s28} converges quickly because
the $\sigma_n$ decrease rapidly with $n$, so typically only $\sigma_n$
for $n \le 5$ are needed to get accurate results for $E$.

The expression in Eq.~\eqref{s28} also allows us to determine under
what conditions the relative error $E$ becomes minimal.  For this to
occur we need $\sigma_1 \ne 0$, since otherwise $E$ takes its maximum
value of 1.  The sum within the brackets in Eq.~\eqref{s28} is
composed of only non-negative terms, and $E$ is smallest when this sum
is minimal.  This can be achieved by setting $\sigma_n = 0$ for all
$n\ge 2$.  Thus $E$ is bounded from below by
\begin{equation}\label{s29}
E \ge 1 - \frac{\bar{I}\gamma_O^2 \sigma_1^2}{(\gamma_I+\gamma_O)^2}\left[\gamma_O \sigma_0 + \sigma_1^2 \frac{\gamma_O \bar{I}}{\gamma_O+\gamma_I} \right]^{-1},
\end{equation}
where the equality is only reached when $R(I)$ has an optimal linear
form, $R_\text{opt}(I) = \sigma_0 v_0(I) + \sigma_1 v_1(I) = \sigma_0
+ \sigma_1 (I-\bar{I})$.  The right-hand side of Eq.~\eqref{s29} is
minimized with respect to $\gamma_O$ when $\gamma_O = \gamma_I
\sqrt{1+\tilde \Lambda}$, with $\tilde \Lambda \equiv \bar{I} \sigma_1^2/\sigma_0
\gamma_I$.  At this optimal $\gamma_O$, the inequality in
Eq.~\eqref{s29} becomes
\begin{equation}\label{s30}
E \ge \frac{2}{1+\sqrt{1+\tilde \Lambda}} \equiv E_\text{opt}.
\end{equation}

\section{Mapping the enzymatic push-pull loop onto the WK filter}\label{a:map}

The full set of reactions for the enzymatic push-pull loop is given by
\begin{equation}\label{m1}
\begin{split}
\varnothing &\xrightleftharpoons[\gamma_K]{F} K\\
K + S &\xrightleftharpoons[\kappa_\text{u}]{\kappa_\text{b}} S_K
\xrightarrow{\kappa_\text{r}} K + S^\ast,\\
P + S^\ast
&\xrightleftharpoons[\rho_\text{u}]{\rho_\text{b}} S^\ast_P
\xrightarrow{\rho_\text{r}} P + S.
\end{split}
\end{equation}
The corresponding steady-states populations are
\begin{equation}\label{m2}
\begin{split}
\bar{K} &= \frac{F}{\gamma_K}, \quad \bar{S}_K = \frac{F \kappa_+}{\gamma_K \kappa_-}, \quad \bar{S}^\ast = \frac{F \kappa_\text{r} \kappa_+ \rho_-}{\gamma_K \kappa_- \rho_\text{r} \rho_+},\\
\bar{S}^\ast_P &= \frac{F \kappa_\text{r} \kappa_+}{\gamma_K \kappa_- \rho_\text{r}},
\end{split}
\end{equation}
where $\kappa_+ = \kappa_\text{b} \bar{S}$, $\kappa_- =
\kappa_\text{u}+ \kappa_\text{r}$, $\kappa = \kappa_+ + \kappa_-$,
$\rho_+ = \rho_\text{b} \bar{P}$, $\rho_- = \rho_\text{u}+
\rho_\text{r}$, and $\rho = \rho_+ + \rho_-$.  

For the system in Eq.~\eqref{m1}, the associated set of chemical
Langevin equations is
\begin{equation}\label{m3}
\begin{split}
\frac{dK}{dt} &= F - \gamma_K K - \kappa_\text{b} KS +(\kappa_\text{u}+\kappa_\text{r})S_K + n_1 + n_2\\
& \qquad+n_3,\\
\frac{dS_K}{dt} &= \kappa_\text{b} KS -(\kappa_\text{u}+\kappa_\text{r})S_K -n_2 -n_3,\\
\frac{dS^\ast}{dt} &= \kappa_\text{r} S_K -\rho_\text{b} P S^\ast + \rho_\text{u} S^\ast_P + n_3 +n_4,\\
\frac{dS^\ast_P}{dt} &= \rho_\text{b} P S^\ast - (\rho_\text{u} + \rho_\text{r})S^\ast_P -n_4 + n_5,\\
\frac{dP}{dt} &= -\frac{dS^\ast_P}{dt}, \qquad \frac{dS}{dt} = -\frac{dS_K}{dt} - \frac{dS^\ast}{dt} - \frac{dS^\ast_P}{dt},
\end{split}
\end{equation}
where the equations on the last line come from the assumptions that
the total populations of free/bound phosphatase ($P+S^\ast_P$) and
free/bound substrate in all its forms ($S+S_K+S^\ast+S^\ast_P$) remain
constant.  The noise terms $n_i(t) = \sqrt{P_{n_i}}\, \eta_i(t)$,
where the $\eta_i(t)$ are Gaussian white noise functions with
correlations $\langle \eta_i(t) \eta_j(t^\prime)\rangle = \delta_{ij}
\delta(t-t^\prime)$.  The constants $P_{n_i}$ are the power spectra of
the noise terms, given by
\begin{equation}\label{m4}
\begin{split}
P_1 &= 2\gamma_K \bar{K}, \quad P_2 = \kappa_\text{b} \bar{K} \bar{S} + \kappa_\text{u} \bar{S}_K, \quad P_3 = \kappa_\text{r} \bar{S}_K,\\
P_4 &= \rho_\text{b} \bar{P}\bar{S}^\ast + \rho_\text{u} \bar{S}^\ast_P, \quad P_5 = \rho_\text{r} \bar{S}^\ast_P.
\end{split}
\end{equation}

We are interested in how the kinase input signal $\delta I = \delta K
+ \delta S_K$ is transduced into the active substrate output $\delta O =
\delta S^\ast + \delta S^\ast_P$, and in particular whether the system
can be approximately mapped onto a WK noise filter of the form given
in the main text (Eq.~2).  (Recall that $\delta x(t) \equiv x(t) -
\bar{x}$ for any time series $x(t)$.)  Since the WK description hinges
on the form of the correlation functions of input and output, we will
need to calculate such correlations for the dynamical equations in
Eq.~\eqref{m3}.  After linearizing these equations, it will be easier
to work in Fourier space, where the Fourier-transformed correlation
functions correspond to power spectra: $P_{\delta x}(\omega) = \int dt
\langle \delta x(t) \delta x(0) \rangle e^{i\omega t}$ for a given
$\delta x(t)$.  Hence it will useful, before proceeding further, to
recast main text Eq.~2, the time-domain noise filter, as a
Fourier-space relation in terms of the power spectra.  The result is
\begin{equation}\label{m5}
\begin{split}
P_{\delta I}(\omega) &= \frac{2 F \gamma_I^{-2}}{1+ (\omega/\gamma_I)^2},\\
P_{\delta O}(\omega) &= \frac{(R_1/\gamma_O G)^2}{1+ (\omega/\gamma_O)^2} \left[G^2 P_{\delta I}(\omega) + \frac{2F(G/\gamma_I)^{2}}{\Lambda}\right].
\end{split}
\end{equation}
Our goal in this section is to show that $P_{\delta I}$ and $P_{\delta
  O}$ calculated for the enzymatic push-pull loop in Eq.~\eqref{m3}
have the approximate form of Eq.~\eqref{m5}, with effective values for
$\gamma_I$, $\gamma_O$, $R_1$, and $\Lambda$ expressed in terms of the
loop reaction rate parameters.

The equilibrium populations $\bar{K}$ and $\bar{S}_K$ scale with
$\bar{I}$ as $\bar{K} = (\kappa_-/\kappa) \bar{I}$ and $\bar{S}_K =
(\kappa_+/\kappa) \bar{I}$.  Similarly, $\bar{S}^\ast = (\rho_-/\rho)
\bar{O}$ and $\bar{S}^\ast_P = (\rho_+/\rho) \bar{O}$.  Each deviation
from the mean---$\delta K$, $\delta S_K$, $\delta S^\ast$, and $\delta
S^\ast_P$---we will explicitly divide into a component that scales
with $\delta I$ or $\delta O$ like the mean population (the ``slowly''
varying component), and the remainder (the ``quickly'' varying
component, denoted with subscript $q$):
\begin{equation}\label{m6}
\begin{split}
\delta K &= \frac{\kappa_-}{\kappa} \delta I + \delta I_q, \quad \delta S_K = \frac{\kappa_+}{\kappa} \delta I - \delta I_q,\\
\delta S^\ast &= \frac{\rho_-}{\rho} \delta O + \delta O_q, \quad \delta S^\ast_P = \frac{\rho_+}{\rho} \delta O - \delta O_q.
\end{split}
\end{equation}
We can interpret Eq.~\eqref{m6} as defining a change of variables from
the set $\delta K$, $\delta S_K$, $\delta S^\ast$, and $\delta
S^\ast_P$ to the set $\delta O$, $\delta O_q$, $\delta I$, $\delta
I_q$.  The nomenclature of slow and quick components comes from the
fact that if the enzymatic reaction rates ($\kappa_+$, $\kappa_-$,
$\rho_+$, $\rho_-$) are made extremely rapid, the characteristic time
scales for the $\delta I_q$ and $\delta O_q$ fluctuations become so
small that the quick components can be neglected, since there would be
nearly instantaneous equilibration between the free and bound enzyme
populations.  In general, however, we cannot assume this limiting case
always holds, so we will take into account both the slow and quick
components in our analysis.

The dynamical system of Eq.~\eqref{m3}, after linearization, Fourier
transform, and the change of variables in Eq.~\eqref{m6}, takes the
form of linear system of equations that can be written in matrix form
as
\begin{widetext}
\begin{equation}\label{m7}
\begin{pmatrix} -i \omega + \frac{\kappa_-\gamma_K}{\kappa_+} & \gamma_K & 0 & 0\\
\frac{\kappa_+ (i \omega \bar{S} -\kappa_+ \bar{K})}{\kappa \bar{S}} & -i\omega + \kappa + \frac{\kappa_+ \bar{K}}{\bar{S}} & - \frac{\kappa_+ \bar{K}}{\bar{S}} & 0\\
-\frac{\kappa_\text{r} \kappa_+}{\kappa} & \kappa_\text{r} & -i\omega + \frac{\rho_\text{r} \rho_+}{\rho} & -\rho_\text{r}\\
0 & 0 & \frac{\rho_+(i\omega \bar{P}-\rho_+\bar{S}^\ast)}{\rho \bar{P}}& -i \omega + \rho + \frac{\rho_+\bar{S}^\ast}{\bar{P}}
\end{pmatrix}
\begin{pmatrix} \delta \tilde I \\ \delta \tilde I_q \\ \delta \tilde O \\ \delta \tilde O_q \end{pmatrix} = \begin{pmatrix}\tilde n_1\\ \tilde n_2+\tilde n_3\\ \tilde n_3+\tilde n_5\\\tilde n_4-\tilde n_5 \end{pmatrix},
\end{equation}
\end{widetext}
where $\tilde x(\omega)$ denotes the Fourier transform of $x(t)$.
Eq.~\eqref{m7} can be solved analytically for $\delta \tilde O$,
$\delta \tilde O_q$, $\delta \tilde I$, $\delta \tilde I_q$, though
for simplicity we will not write out the full solutions, since these
would take up too much space.  Rather we will sketch out the basic
approach to calculating and approximating the associated power
spectra.  The structure of the solutions to Eq.~\eqref{m7}, for
example $\delta \tilde I$, is a linear combination of the the noise
functions, $\delta \tilde I(\omega) = \sum_{i=1}^5 a_{\delta
  I,i}(\omega) \tilde n_i(\omega)$, with coefficients $a_{\delta
  I,i}(\omega)$.  The corresponding power spectrum is $P_{\delta
  I}(\omega) = \sum_{i=1}^5 |a_{\delta I,i}(\omega)|^2 P_{n_i}$, with
$P_{n_i}$ given by Eq.~\eqref{m4}.  The function $P_{\delta
  I}(\omega)$ can be written out in the form of a rational function
with even powers of $\omega$ in the numerator and denominator,
\begin{equation}\label{m8}
P_{\delta I}(\omega) = \frac{\sum_{i=0}^N n_{\delta I,i} \omega^{2i}}{1+\sum_{i=1}^D d_{\delta I,i}\omega^{2i}},
\end{equation}
where $n_{\delta I,i}$ and $d_{\delta I,i}$ are coefficients
independent of $\omega$, and $N=3$, $D=4$ for the case of $P_{\delta
  I}$.  In order to simplify Eq.~\eqref{m8} further, we will make two
assumptions: (i) The characteristic time scale over which the input
signal varies, $\gamma_K^{-1}$, is much longer than the characteristic
time scales of the enzymatic reactions, $\kappa^{-1}_\alpha$ and
$\rho^{-1}_\alpha$, where $\alpha$ denotes the various subscripts $+$,
$-$, and $\text{r}$.  For the parameters in the main text,
$\gamma_K^{-1} \sim {\cal O}(10^2\:\:\text{s})$, while
$\kappa^{-1}_\alpha$, $\rho^{-1}_\alpha \sim {\cal O}(10^{-1} -
10^0\:\:\text{s})$.  This the physically interesting regime, since we
can expect the system to efficiently transduce signals that vary more
slowly than the intrinsic reactions that carry out the transduction.
Limiting our focus to frequencies $\omega \ll \kappa^{-1}_\alpha$,
$\rho^{-1}_\alpha$, it turns out that the higher order powers of
$\omega$ in both the numerator and denominator of Eq.~\eqref{m8} are
negligible, and the power spectrum can be approximated by
\begin{equation}\label{m9}
P_{\delta I}(\omega) \approx \frac{n_{\delta I,0}}{1+d_{\delta I,1} \omega^2}.
\end{equation}
(ii) We assume that the system is in the regime where $\bar{K} =
F/\gamma_K \ll \bar{S}, \bar{P}$.  Thus we will expand the
coefficients $n_{\delta I,0}$ and $d_{\delta I,1}$ in Eq.~\eqref{m9}
up to first order in $\bar{K}/\bar{S}$ and $\bar{K}/\bar{P}$,
resulting in a $P_{\delta I}(\omega)$ that has the form of
Eq.~\eqref{m5}.  Namely, $n_{\delta I,0} \approx 2F\gamma_I^{-2}$ and
$d_{\delta I,1} \approx \gamma_I^{-2}$, where the effective $\gamma_I^{-2}$
is given by
\begin{equation}\label{m10}
\gamma_I = \frac{\kappa_- \gamma_K}{\kappa} + \frac{\kappa_+^2 (\rho_\text{r} \rho_+ + \kappa_\text{r} \rho)\gamma_K \bar{K}}{\kappa^2 \rho_\text{r} \rho_+\bar{S}}.
\end{equation}

In an analogous manner we can find the correspondence between
$P_{\delta O}(\omega)$ and the form in Eq.~\eqref{m5}, leading to the
following expressions for the remaining effective parameters,
\begin{widetext}
\begin{equation}\label{m11}
\begin{split}
\gamma_O &= \frac{\rho_\text{r} \rho_+}{\sqrt{\rho^2 - 2 \rho_\text{r} \rho_+}} - \frac{\kappa_\text{r} \kappa_+ \rho_+ \rho_- \rho \bar{K}}{\kappa_-(\rho^2 - 2 \rho_\text{r} \rho_+)^{3/2}\bar{P}},\qquad
R_1 = \frac{\kappa_\text{r} \kappa_+ \rho}{\kappa \sqrt{\rho^2 - 2 \rho_\text{r} \rho_+}} - \frac{2\kappa_\text{r}^2 \kappa_+^2 \rho_+ \rho_- \bar{K}}{\kappa_-\kappa (\rho^2 - 2 \rho_\text{r} \rho_+)^{3/2}\bar{P}},\\
\Lambda &= \frac{\kappa_\text{r} \kappa_+ \kappa^2 \rho^2}{\gamma_K \kappa_- ( \rho^2 (\kappa^2 - \kappa_\text{r} \kappa_+) - \rho_\text{r} \rho_+ \kappa^2)} + \frac{\kappa_\text{r} \kappa_+^2 \kappa \rho \bar{K} }{\gamma_K \rho_\text{r} \kappa_-^2 \rho_+ \left(\rho^2 \left(\kappa_\text{r} \kappa_+-\kappa^2\right)+\rho_\text{r} \kappa^2 \rho_+\right)^2 \bar{S} \bar{P}}\left[\left\{\kappa_\text{r}^2 \kappa_+ \rho^3 (\kappa_+-\kappa_-)\right.\right.\\
&\qquad\qquad\left.\left.+\kappa_\text{r} \rho \left(\rho_\text{r} \rho_+ \left(2 \kappa_+^3+\kappa_+^2 (2 \kappa_-+\rho)-2 \kappa_+ \kappa_-^2+\kappa_-^2 (\rho-2 \kappa_-)\right) -2 \kappa^2 \rho^2 (\kappa_+-\kappa_-)\right)\right.\right.\\
&\qquad\qquad\left.\left. +2 \rho_\text{r} \kappa_+ \kappa^2 \rho_+ \left(\rho_\text{r} \rho_+-\rho^2\right)\right\} \rho \bar{P} +\kappa_\text{r} \rho_\text{r} \kappa^3 \rho_+ \rho_- (\rho_--\rho_+)\bar{S} \right].
\end{split}
\end{equation}
\end{widetext}
The results in Eqs.~\eqref{m10} and \eqref{m11}, without the
first-order corrections in $\bar{K}/\bar{S}$ and $\bar{K}/\bar{P}$,
correspond to main text Eq.~(8).


\begin{thebibliography}{31}%
\makeatletter
\providecommand \@ifxundefined [1]{%
 \@ifx{#1\undefined}
}%
\providecommand \@ifnum [1]{%
 \ifnum #1\expandafter \@firstoftwo
 \else \expandafter \@secondoftwo
 \fi
}%
\providecommand \@ifx [1]{%
 \ifx #1\expandafter \@firstoftwo
 \else \expandafter \@secondoftwo
 \fi
}%
\providecommand \natexlab [1]{#1}%
\providecommand \enquote  [1]{``#1''}%
\providecommand \bibnamefont  [1]{#1}%
\providecommand \bibfnamefont [1]{#1}%
\providecommand \citenamefont [1]{#1}%
\providecommand \href@noop [0]{\@secondoftwo}%
\providecommand \href [0]{\begingroup \@sanitize@url \@href}%
\providecommand \@href[1]{\@@startlink{#1}\@@href}%
\providecommand \@@href[1]{\endgroup#1\@@endlink}%
\providecommand \@sanitize@url [0]{\catcode `\\12\catcode `\$12\catcode
  `\&12\catcode `\#12\catcode `\^12\catcode `\_12\catcode `\%12\relax}%
\providecommand \@@startlink[1]{}%
\providecommand \@@endlink[0]{}%
\providecommand \url  [0]{\begingroup\@sanitize@url \@url }%
\providecommand \@url [1]{\endgroup\@href {#1}{\urlprefix }}%
\providecommand \urlprefix  [0]{URL }%
\providecommand \Eprint [0]{\href }%
\providecommand \doibase [0]{http://dx.doi.org/}%
\providecommand \selectlanguage [0]{\@gobble}%
\providecommand \bibinfo  [0]{\@secondoftwo}%
\providecommand \bibfield  [0]{\@secondoftwo}%
\providecommand \translation [1]{[#1]}%
\providecommand \BibitemOpen [0]{}%
\providecommand \bibitemStop [0]{}%
\providecommand \bibitemNoStop [0]{.\EOS\space}%
\providecommand \EOS [0]{\spacefactor3000\relax}%
\providecommand \BibitemShut  [1]{\csname bibitem#1\endcsname}%
\let\auto@bib@innerbib\@empty
%</preamble>
\bibitem [{\citenamefont {Altschuler}\ and\ \citenamefont
  {Wu}(2010)}]{Altschuler2010}%
  \BibitemOpen
  \bibfield  {author} {\bibinfo {author} {\bibfnamefont {Steven~J}\
  \bibnamefont {Altschuler}}\ and\ \bibinfo {author} {\bibfnamefont {Lani~F}\
  \bibnamefont {Wu}},\ }\bibfield  {title} {\enquote {\bibinfo {title}
  {Cellular heterogeneity: do differences make a difference?}}\ }\href@noop {}
  {\bibfield  {journal} {\bibinfo  {journal} {Cell}\ }\textbf {\bibinfo
  {volume} {141}},\ \bibinfo {pages} {559--563} (\bibinfo {year}
  {2010})}\BibitemShut {NoStop}%
\bibitem [{\citenamefont {Cai}\ \emph {et~al.}(2008)\citenamefont {Cai},
  \citenamefont {Dalal},\ and\ \citenamefont {Elowitz}}]{Cai08}%
  \BibitemOpen
  \bibfield  {author} {\bibinfo {author} {\bibfnamefont {L.}~\bibnamefont
  {Cai}}, \bibinfo {author} {\bibfnamefont {C.~K.}\ \bibnamefont {Dalal}}, \
  and\ \bibinfo {author} {\bibfnamefont {M.~B.}\ \bibnamefont {Elowitz}},\
  }\bibfield  {title} {\enquote {\bibinfo {title} {Frequency-modulated nuclear
  localization bursts coordinate gene regulation},}\ }\href@noop {} {\bibfield
  {journal} {\bibinfo  {journal} {Nature}\ }\textbf {\bibinfo {volume} {455}},\
  \bibinfo {pages} {485--U16} (\bibinfo {year} {2008})}\BibitemShut {NoStop}%
\bibitem [{\citenamefont {Wiener}(1949)}]{Wiener49}%
  \BibitemOpen
  \bibfield  {author} {\bibinfo {author} {\bibfnamefont {N.}~\bibnamefont
  {Wiener}},\ }\href@noop {} {\emph {\bibinfo {title} {Extrapolation,
  Interpolation and Smoothing of Stationary Times Series}}}\ (\bibinfo
  {publisher} {Wiley},\ \bibinfo {address} {New York},\ \bibinfo {year}
  {1949})\BibitemShut {NoStop}%
\bibitem [{\citenamefont {Kolmogorov}(1941)}]{Kolmogorov41}%
  \BibitemOpen
  \bibfield  {author} {\bibinfo {author} {\bibfnamefont {A.~N.}\ \bibnamefont
  {Kolmogorov}},\ }\bibfield  {title} {\enquote {\bibinfo {title}
  {Interpolation and extrapolation of stationary random sequences},}\
  }\href@noop {} {\bibfield  {journal} {\bibinfo  {journal} {Izv. Akad. Nauk
  SSSR., Ser. Mat.}\ }\textbf {\bibinfo {volume} {5}},\ \bibinfo {pages}
  {3--14} (\bibinfo {year} {1941})}\BibitemShut {NoStop}%
\bibitem [{\citenamefont {Mettetal}\ \emph {et~al.}(2008)\citenamefont
  {Mettetal}, \citenamefont {Muzzey}, \citenamefont {Gomez-Uribe},\ and\
  \citenamefont {van Oudenaarden}}]{Mettetal08}%
  \BibitemOpen
  \bibfield  {author} {\bibinfo {author} {\bibfnamefont {J.~T.}\ \bibnamefont
  {Mettetal}}, \bibinfo {author} {\bibfnamefont {D.}~\bibnamefont {Muzzey}},
  \bibinfo {author} {\bibfnamefont {C.}~\bibnamefont {Gomez-Uribe}}, \ and\
  \bibinfo {author} {\bibfnamefont {A.}~\bibnamefont {van Oudenaarden}},\
  }\bibfield  {title} {\enquote {\bibinfo {title} {The frequency dependence of
  osmo-adaptation in {Saccharomyces} cerevisiae},}\ }\href@noop {} {\bibfield
  {journal} {\bibinfo  {journal} {Science}\ }\textbf {\bibinfo {volume}
  {319}},\ \bibinfo {pages} {482--484} (\bibinfo {year} {2008})}\BibitemShut
  {NoStop}%
\bibitem [{\citenamefont {Hersen}\ \emph {et~al.}(2008)\citenamefont {Hersen},
  \citenamefont {McClean}, \citenamefont {Mahadevan},\ and\ \citenamefont
  {Ramanathan}}]{Hersen08}%
  \BibitemOpen
  \bibfield  {author} {\bibinfo {author} {\bibfnamefont {P.}~\bibnamefont
  {Hersen}}, \bibinfo {author} {\bibfnamefont {M.~N.}\ \bibnamefont {McClean}},
  \bibinfo {author} {\bibfnamefont {L.}~\bibnamefont {Mahadevan}}, \ and\
  \bibinfo {author} {\bibfnamefont {S.}~\bibnamefont {Ramanathan}},\ }\bibfield
   {title} {\enquote {\bibinfo {title} {Signal processing by the {HOG} {MAP}
  kinase pathway},}\ }\href@noop {} {\bibfield  {journal} {\bibinfo  {journal}
  {Proc. Natl. Acad. Sci. U.S.A.}\ }\textbf {\bibinfo {volume} {105}},\
  \bibinfo {pages} {7165--7170} (\bibinfo {year} {2008})}\BibitemShut {NoStop}%
\bibitem [{\citenamefont {Cheong}\ \emph {et~al.}(2011)\citenamefont {Cheong},
  \citenamefont {Rhee}, \citenamefont {Wang}, \citenamefont {Nemenman},\ and\
  \citenamefont {Levchenko}}]{Cheong11}%
  \BibitemOpen
  \bibfield  {author} {\bibinfo {author} {\bibfnamefont {R.}~\bibnamefont
  {Cheong}}, \bibinfo {author} {\bibfnamefont {A.}~\bibnamefont {Rhee}},
  \bibinfo {author} {\bibfnamefont {C.~J.}\ \bibnamefont {Wang}}, \bibinfo
  {author} {\bibfnamefont {I.}~\bibnamefont {Nemenman}}, \ and\ \bibinfo
  {author} {\bibfnamefont {A.}~\bibnamefont {Levchenko}},\ }\bibfield  {title}
  {\enquote {\bibinfo {title} {Information {Transduction} {Capacity} of {Noisy}
  {Biochemical} {Signaling} networks},}\ }\href@noop {} {\bibfield  {journal}
  {\bibinfo  {journal} {Science}\ }\textbf {\bibinfo {volume} {334}},\ \bibinfo
  {pages} {354--358} (\bibinfo {year} {2011})}\BibitemShut {NoStop}%
\bibitem [{\citenamefont {Balazsi}\ \emph {et~al.}(2011)\citenamefont
  {Balazsi}, \citenamefont {van Oudenaarden},\ and\ \citenamefont
  {Collins}}]{Balazsi11Cell}%
  \BibitemOpen
  \bibfield  {author} {\bibinfo {author} {\bibfnamefont {Gabor}\ \bibnamefont
  {Balazsi}}, \bibinfo {author} {\bibfnamefont {Alexander}\ \bibnamefont {van
  Oudenaarden}}, \ and\ \bibinfo {author} {\bibfnamefont {James~J.}\
  \bibnamefont {Collins}},\ }\bibfield  {title} {\enquote {\bibinfo {title}
  {{Cellular Decision Making and Biological Noise: From Microbes to
  Mammals}},}\ }\href@noop {} {\bibfield  {journal} {\bibinfo  {journal}
  {{Cell}}\ }\textbf {\bibinfo {volume} {144}},\ \bibinfo {pages} {910--925}
  (\bibinfo {year} {2011})}\BibitemShut {NoStop}%
\bibitem [{\citenamefont {Bowsher}\ \emph {et~al.}(2013)\citenamefont
  {Bowsher}, \citenamefont {Voliotis},\ and\ \citenamefont
  {Swain}}]{Bowsher13}%
  \BibitemOpen
  \bibfield  {author} {\bibinfo {author} {\bibfnamefont {Clive~G}\ \bibnamefont
  {Bowsher}}, \bibinfo {author} {\bibfnamefont {Margaritis}\ \bibnamefont
  {Voliotis}}, \ and\ \bibinfo {author} {\bibfnamefont {Peter~S}\ \bibnamefont
  {Swain}},\ }\bibfield  {title} {\enquote {\bibinfo {title} {The fidelity of
  dynamic signaling by noisy biomolecular networks},}\ }\href@noop {}
  {\bibfield  {journal} {\bibinfo  {journal} {PLoS Comp. Biol.}\ }\textbf
  {\bibinfo {volume} {9}},\ \bibinfo {pages} {e1002965} (\bibinfo {year}
  {2013})}\BibitemShut {NoStop}%
\bibitem [{\citenamefont {Thattai}\ and\ \citenamefont {van
  Oudenaarden}(2002)}]{Thattai02}%
  \BibitemOpen
  \bibfield  {author} {\bibinfo {author} {\bibfnamefont {M.}~\bibnamefont
  {Thattai}}\ and\ \bibinfo {author} {\bibfnamefont {A.}~\bibnamefont {van
  Oudenaarden}},\ }\bibfield  {title} {\enquote {\bibinfo {title} {Attenuation
  of noise in ultrasensitive signaling cascades},}\ }\href@noop {} {\bibfield
  {journal} {\bibinfo  {journal} {Biophys. J.}\ }\textbf {\bibinfo {volume}
  {82}},\ \bibinfo {pages} {2943--2950} (\bibinfo {year} {2002})}\BibitemShut
  {NoStop}%
\bibitem [{\citenamefont {T\u{a}nase-Nicola}\ \emph {et~al.}(2006)\citenamefont
  {T\u{a}nase-Nicola}, \citenamefont {Warren},\ and\ \citenamefont {ten
  Wolde}}]{TanaseNicola06}%
  \BibitemOpen
  \bibfield  {author} {\bibinfo {author} {\bibfnamefont {S.}~\bibnamefont
  {T\u{a}nase-Nicola}}, \bibinfo {author} {\bibfnamefont {P.~B.}\ \bibnamefont
  {Warren}}, \ and\ \bibinfo {author} {\bibfnamefont {P.~R.}\ \bibnamefont {ten
  Wolde}},\ }\bibfield  {title} {\enquote {\bibinfo {title} {Signal detection,
  modularity, and the correlation between extrinsic and intrinsic noise in
  biochemical networks},}\ }\href@noop {} {\bibfield  {journal} {\bibinfo
  {journal} {Phys. Rev. Lett.}\ }\textbf {\bibinfo {volume} {97}},\ \bibinfo
  {pages} {068102} (\bibinfo {year} {2006})}\BibitemShut {NoStop}%
\bibitem [{\citenamefont {Stadtman}\ and\ \citenamefont
  {Chock}(1977)}]{Stadtman77}%
  \BibitemOpen
  \bibfield  {author} {\bibinfo {author} {\bibfnamefont {E.~R.}\ \bibnamefont
  {Stadtman}}\ and\ \bibinfo {author} {\bibfnamefont {P.~B.}\ \bibnamefont
  {Chock}},\ }\bibfield  {title} {\enquote {\bibinfo {title} {Superiority of
  interconvertible enzyme cascades in metabolic-regulation - analysis of
  monocyclic systems},}\ }\href@noop {} {\bibfield  {journal} {\bibinfo
  {journal} {Proc. Natl. Acad. Sci. U.S.A.}\ }\textbf {\bibinfo {volume}
  {74}},\ \bibinfo {pages} {2761--2765} (\bibinfo {year} {1977})}\BibitemShut
  {NoStop}%
\bibitem [{\citenamefont {Goldbeter}\ and\ \citenamefont
  {Koshland}(1981)}]{Goldbeter81}%
  \BibitemOpen
  \bibfield  {author} {\bibinfo {author} {\bibfnamefont {A.}~\bibnamefont
  {Goldbeter}}\ and\ \bibinfo {author} {\bibfnamefont {D.~E.}\ \bibnamefont
  {Koshland}},\ }\bibfield  {title} {\enquote {\bibinfo {title} {An amplified
  sensitivity arising from covalent modification in biological-systems},}\
  }\href@noop {} {\bibfield  {journal} {\bibinfo  {journal} {Proc. Natl. Acad.
  Sci. U.S.A.}\ }\textbf {\bibinfo {volume} {78}},\ \bibinfo {pages}
  {6840--6844} (\bibinfo {year} {1981})}\BibitemShut {NoStop}%
\bibitem [{\citenamefont {Detwiler}\ \emph {et~al.}(2000)\citenamefont
  {Detwiler}, \citenamefont {Ramanathan}, \citenamefont {Sengupta},\ and\
  \citenamefont {Shraiman}}]{Detwiler00}%
  \BibitemOpen
  \bibfield  {author} {\bibinfo {author} {\bibfnamefont {P.~B.}\ \bibnamefont
  {Detwiler}}, \bibinfo {author} {\bibfnamefont {S.}~\bibnamefont
  {Ramanathan}}, \bibinfo {author} {\bibfnamefont {A.}~\bibnamefont
  {Sengupta}}, \ and\ \bibinfo {author} {\bibfnamefont {B.~I.}\ \bibnamefont
  {Shraiman}},\ }\bibfield  {title} {\enquote {\bibinfo {title} {Engineering
  aspects of enzymatic signal transduction: {Photoreceptors} in the retina},}\
  }\href@noop {} {\bibfield  {journal} {\bibinfo  {journal} {Biophys. J.}\
  }\textbf {\bibinfo {volume} {79}},\ \bibinfo {pages} {2801--2817} (\bibinfo
  {year} {2000})}\BibitemShut {NoStop}%
\bibitem [{\citenamefont {Heinrich}\ \emph {et~al.}(2002)\citenamefont
  {Heinrich}, \citenamefont {Neel},\ and\ \citenamefont
  {Rapoport}}]{Heinrich2002}%
  \BibitemOpen
  \bibfield  {author} {\bibinfo {author} {\bibfnamefont {Reinhart}\
  \bibnamefont {Heinrich}}, \bibinfo {author} {\bibfnamefont {Benjamin~G}\
  \bibnamefont {Neel}}, \ and\ \bibinfo {author} {\bibfnamefont {Tom~A}\
  \bibnamefont {Rapoport}},\ }\bibfield  {title} {\enquote {\bibinfo {title}
  {Mathematical models of protein kinase signal transduction},}\ }\href@noop {}
  {\bibfield  {journal} {\bibinfo  {journal} {Molecular cell}\ }\textbf
  {\bibinfo {volume} {9}},\ \bibinfo {pages} {957--970} (\bibinfo {year}
  {2002})}\BibitemShut {NoStop}%
\bibitem [{\citenamefont {Samoilov}\ \emph {et~al.}(2005)\citenamefont
  {Samoilov}, \citenamefont {Plyasunov},\ and\ \citenamefont
  {Arkin}}]{Samoilov05}%
  \BibitemOpen
  \bibfield  {author} {\bibinfo {author} {\bibfnamefont {M.}~\bibnamefont
  {Samoilov}}, \bibinfo {author} {\bibfnamefont {S.}~\bibnamefont {Plyasunov}},
  \ and\ \bibinfo {author} {\bibfnamefont {A.~P.}\ \bibnamefont {Arkin}},\
  }\bibfield  {title} {\enquote {\bibinfo {title} {Stochastic amplification and
  signaling in enzymatic futile cycles through noise-induced bistability with
  oscillations},}\ }\href@noop {} {\bibfield  {journal} {\bibinfo  {journal}
  {Proc. Natl. Acad. Sci. U.S.A.}\ }\textbf {\bibinfo {volume} {102}},\
  \bibinfo {pages} {2310--2315} (\bibinfo {year} {2005})}\BibitemShut {NoStop}%
\bibitem [{\citenamefont {Levine}\ \emph {et~al.}(2007)\citenamefont {Levine},
  \citenamefont {Kueh},\ and\ \citenamefont {Mirny}}]{Levine07}%
  \BibitemOpen
  \bibfield  {author} {\bibinfo {author} {\bibfnamefont {J.}~\bibnamefont
  {Levine}}, \bibinfo {author} {\bibfnamefont {H.~Y.}\ \bibnamefont {Kueh}}, \
  and\ \bibinfo {author} {\bibfnamefont {L.}~\bibnamefont {Mirny}},\ }\bibfield
   {title} {\enquote {\bibinfo {title} {Intrinsic fluctuations, robustness, and
  tunability in signaling cycles},}\ }\href@noop {} {\bibfield  {journal}
  {\bibinfo  {journal} {Biophys. J.}\ }\textbf {\bibinfo {volume} {92}},\
  \bibinfo {pages} {4473--4481} (\bibinfo {year} {2007})}\BibitemShut {NoStop}%
\bibitem [{\citenamefont {Gomez-Uribe}\ \emph {et~al.}(2007)\citenamefont
  {Gomez-Uribe}, \citenamefont {Verghese},\ and\ \citenamefont
  {Mirny}}]{GomezUribe07}%
  \BibitemOpen
  \bibfield  {author} {\bibinfo {author} {\bibfnamefont {C.}~\bibnamefont
  {Gomez-Uribe}}, \bibinfo {author} {\bibfnamefont {G.~C.}\ \bibnamefont
  {Verghese}}, \ and\ \bibinfo {author} {\bibfnamefont {L.~A.}\ \bibnamefont
  {Mirny}},\ }\bibfield  {title} {\enquote {\bibinfo {title} {Operating regimes
  of signaling cycles: {Statics,} dynamics, and noise filtering},}\ }\href@noop
  {} {\bibfield  {journal} {\bibinfo  {journal} {PLoS Comput. Biol.}\ }\textbf
  {\bibinfo {volume} {3}},\ \bibinfo {pages} {2487--2497} (\bibinfo {year}
  {2007})}\BibitemShut {NoStop}%
\bibitem [{\citenamefont {Lestas}\ \emph {et~al.}(2010)\citenamefont {Lestas},
  \citenamefont {Vinnicombe},\ and\ \citenamefont {Paulsson}}]{Lestas10}%
  \BibitemOpen
  \bibfield  {author} {\bibinfo {author} {\bibfnamefont {I.}~\bibnamefont
  {Lestas}}, \bibinfo {author} {\bibfnamefont {G.}~\bibnamefont {Vinnicombe}},
  \ and\ \bibinfo {author} {\bibfnamefont {J.}~\bibnamefont {Paulsson}},\
  }\bibfield  {title} {\enquote {\bibinfo {title} {Fundamental limits on the
  suppression of molecular fluctuations},}\ }\href@noop {} {\bibfield
  {journal} {\bibinfo  {journal} {Nature}\ }\textbf {\bibinfo {volume} {467}},\
  \bibinfo {pages} {174--178} (\bibinfo {year} {2010})}\BibitemShut {NoStop}%
\bibitem [{\citenamefont {Roman}(2005)}]{Roman}%
  \BibitemOpen
  \bibfield  {author} {\bibinfo {author} {\bibfnamefont {S}~\bibnamefont
  {Roman}},\ }\href@noop {} {\emph {\bibinfo {title} {The Umbral Calculus}}}\
  (\bibinfo  {publisher} {{Dover}},\ \bibinfo {year} {2005})\BibitemShut
  {NoStop}%
\bibitem [{\citenamefont {Mugler}\ \emph {et~al.}(2010)\citenamefont {Mugler},
  \citenamefont {Walczak},\ and\ \citenamefont {Wiggins}}]{Mugler10}%
  \BibitemOpen
  \bibfield  {author} {\bibinfo {author} {\bibfnamefont {A.}~\bibnamefont
  {Mugler}}, \bibinfo {author} {\bibfnamefont {A.~M.}\ \bibnamefont {Walczak}},
  \ and\ \bibinfo {author} {\bibfnamefont {C.~H.}\ \bibnamefont {Wiggins}},\
  }\bibfield  {title} {\enquote {\bibinfo {title} {Information-optimal
  {Transcriptional} {Response} to {Oscillatory} driving},}\ }\href@noop {}
  {\bibfield  {journal} {\bibinfo  {journal} {Phys. Rev. Lett.}\ }\textbf
  {\bibinfo {volume} {105}},\ \bibinfo {pages} {058101} (\bibinfo {year}
  {2010})}\BibitemShut {NoStop}%
\bibitem [{\citenamefont {Gillespie}(2000)}]{Gillespie00}%
  \BibitemOpen
  \bibfield  {author} {\bibinfo {author} {\bibfnamefont {D.~T.}\ \bibnamefont
  {Gillespie}},\ }\bibfield  {title} {\enquote {\bibinfo {title} {The chemical
  {Langevin} equation},}\ }\href@noop {} {\bibfield  {journal} {\bibinfo
  {journal} {J. Chem. Phys.}\ }\textbf {\bibinfo {volume} {113}},\ \bibinfo
  {pages} {297--306} (\bibinfo {year} {2000})}\BibitemShut {NoStop}%
\bibitem [{\citenamefont {Bode}\ and\ \citenamefont {Shannon}(1950)}]{Bode50}%
  \BibitemOpen
  \bibfield  {author} {\bibinfo {author} {\bibfnamefont {H.~W.}\ \bibnamefont
  {Bode}}\ and\ \bibinfo {author} {\bibfnamefont {C.~E.}\ \bibnamefont
  {Shannon}},\ }\bibfield  {title} {\enquote {\bibinfo {title} {A simplified
  derivation of linear least square smoothing and prediction theory},}\
  }\href@noop {} {\bibfield  {journal} {\bibinfo  {journal} {Proc. Inst. Radio.
  Engin.}\ }\textbf {\bibinfo {volume} {38}},\ \bibinfo {pages} {417--425}
  (\bibinfo {year} {1950})}\BibitemShut {NoStop}%
\bibitem [{\citenamefont {Bialek}\ \emph {et~al.}(2001)\citenamefont {Bialek},
  \citenamefont {Nemenman},\ and\ \citenamefont {Tishby}}]{Bialek2001}%
  \BibitemOpen
  \bibfield  {author} {\bibinfo {author} {\bibfnamefont {William}\ \bibnamefont
  {Bialek}}, \bibinfo {author} {\bibfnamefont {Ilya}\ \bibnamefont {Nemenman}},
  \ and\ \bibinfo {author} {\bibfnamefont {Naftali}\ \bibnamefont {Tishby}},\
  }\bibfield  {title} {\enquote {\bibinfo {title} {Predictability, complexity,
  and learning},}\ }\href@noop {} {\bibfield  {journal} {\bibinfo  {journal}
  {Neural Comput.}\ }\textbf {\bibinfo {volume} {13}},\ \bibinfo {pages}
  {2409--2463} (\bibinfo {year} {2001})}\BibitemShut {NoStop}%
\bibitem [{\citenamefont {Mugler}\ \emph {et~al.}(2009)\citenamefont {Mugler},
  \citenamefont {Walczak},\ and\ \citenamefont {Wiggins}}]{Mugler09}%
  \BibitemOpen
  \bibfield  {author} {\bibinfo {author} {\bibfnamefont {Andrew}\ \bibnamefont
  {Mugler}}, \bibinfo {author} {\bibfnamefont {Aleksandra~M}\ \bibnamefont
  {Walczak}}, \ and\ \bibinfo {author} {\bibfnamefont {Chris~H}\ \bibnamefont
  {Wiggins}},\ }\bibfield  {title} {\enquote {\bibinfo {title} {Spectral
  solutions to stochastic models of gene expression with bursts and
  regulation},}\ }\href@noop {} {\bibfield  {journal} {\bibinfo  {journal}
  {Phys. Rev. E}\ }\textbf {\bibinfo {volume} {80}},\ \bibinfo {pages} {041921}
  (\bibinfo {year} {2009})}\BibitemShut {NoStop}%
\bibitem [{\citenamefont {Walczak}\ \emph {et~al.}(2009)\citenamefont
  {Walczak}, \citenamefont {Mugler},\ and\ \citenamefont
  {Wiggins}}]{Walczak09}%
  \BibitemOpen
  \bibfield  {author} {\bibinfo {author} {\bibfnamefont {Aleksandra~M}\
  \bibnamefont {Walczak}}, \bibinfo {author} {\bibfnamefont {Andrew}\
  \bibnamefont {Mugler}}, \ and\ \bibinfo {author} {\bibfnamefont {Chris~H}\
  \bibnamefont {Wiggins}},\ }\bibfield  {title} {\enquote {\bibinfo {title} {A
  stochastic spectral analysis of transcriptional regulatory cascades},}\
  }\href@noop {} {\bibfield  {journal} {\bibinfo  {journal} {Proc. Natl. Acad.
  Sci. U.S.A.}\ }\textbf {\bibinfo {volume} {106}},\ \bibinfo {pages}
  {6529--6534} (\bibinfo {year} {2009})}\BibitemShut {NoStop}%
\bibitem [{\citenamefont {Schoeberl}\ \emph {et~al.}(2002)\citenamefont
  {Schoeberl}, \citenamefont {Eichler-Jonsson}, \citenamefont {Gilles},\ and\
  \citenamefont {Muller}}]{Schoeberl02}%
  \BibitemOpen
  \bibfield  {author} {\bibinfo {author} {\bibfnamefont {B.}~\bibnamefont
  {Schoeberl}}, \bibinfo {author} {\bibfnamefont {C.}~\bibnamefont
  {Eichler-Jonsson}}, \bibinfo {author} {\bibfnamefont {E.~D.}\ \bibnamefont
  {Gilles}}, \ and\ \bibinfo {author} {\bibfnamefont {G.}~\bibnamefont
  {Muller}},\ }\bibfield  {title} {\enquote {\bibinfo {title} {Computational
  modeling of the dynamics of the {MAP} kinase cascade activated by surface and
  internalized {EGF} receptors},}\ }\href@noop {} {\bibfield  {journal}
  {\bibinfo  {journal} {Nat. Biotechnol.}\ }\textbf {\bibinfo {volume} {20}},\
  \bibinfo {pages} {370--375} (\bibinfo {year} {2002})}\BibitemShut {NoStop}%
\bibitem [{\citenamefont {Gillespie}(1977)}]{Gillespie77}%
  \BibitemOpen
  \bibfield  {author} {\bibinfo {author} {\bibfnamefont {D.~T.}\ \bibnamefont
  {Gillespie}},\ }\bibfield  {title} {\enquote {\bibinfo {title} {Exact
  stochastic simulation of coupled chemical-reactions},}\ }\href@noop {}
  {\bibfield  {journal} {\bibinfo  {journal} {J. Phys. Chem.}\ }\textbf
  {\bibinfo {volume} {81}},\ \bibinfo {pages} {2340--2361} (\bibinfo {year}
  {1977})}\BibitemShut {NoStop}%
\bibitem [{\citenamefont {Ghaemmaghami}\ \emph {et~al.}(2003)\citenamefont
  {Ghaemmaghami}, \citenamefont {Huh}, \citenamefont {Bower}, \citenamefont
  {Howson}, \citenamefont {Belle}, \citenamefont {Dephoure}, \citenamefont
  {O'Shea},\ and\ \citenamefont {Weissman}}]{Ghaemmaghami2003}%
  \BibitemOpen
  \bibfield  {author} {\bibinfo {author} {\bibfnamefont {Sina}\ \bibnamefont
  {Ghaemmaghami}}, \bibinfo {author} {\bibfnamefont {Won-Ki}\ \bibnamefont
  {Huh}}, \bibinfo {author} {\bibfnamefont {Kiowa}\ \bibnamefont {Bower}},
  \bibinfo {author} {\bibfnamefont {Russell~W}\ \bibnamefont {Howson}},
  \bibinfo {author} {\bibfnamefont {Archana}\ \bibnamefont {Belle}}, \bibinfo
  {author} {\bibfnamefont {Noah}\ \bibnamefont {Dephoure}}, \bibinfo {author}
  {\bibfnamefont {Erin~K}\ \bibnamefont {O'Shea}}, \ and\ \bibinfo {author}
  {\bibfnamefont {Jonathan~S}\ \bibnamefont {Weissman}},\ }\bibfield  {title}
  {\enquote {\bibinfo {title} {Global analysis of protein expression in
  yeast},}\ }\href@noop {} {\bibfield  {journal} {\bibinfo  {journal} {Nature}\
  }\textbf {\bibinfo {volume} {425}},\ \bibinfo {pages} {737--741} (\bibinfo
  {year} {2003})}\BibitemShut {NoStop}%
\bibitem [{\citenamefont {Hinczewski}\ and\ \citenamefont
  {Thirumalai}(2014)}]{HTpreprint}%
  \BibitemOpen
  \bibfield  {author} {\bibinfo {author} {\bibfnamefont {Michael}\ \bibnamefont
  {Hinczewski}}\ and\ \bibinfo {author} {\bibfnamefont {D}~\bibnamefont
  {Thirumalai}},\ }\href@noop {} {\bibfield  {journal} {\bibinfo  {journal}
  {preprint}\ } (\bibinfo {year} {2014})}\BibitemShut {NoStop}%
\bibitem [{\citenamefont {Gel'fond}(1971)}]{Gelfond}%
  \BibitemOpen
  \bibfield  {author} {\bibinfo {author} {\bibfnamefont {A~O}\ \bibnamefont
  {Gel'fond}},\ }\href@noop {} {\emph {\bibinfo {title} {Calculus of finite
  differences}}}\ (\bibinfo  {publisher} {Hindustan Publ. Corp.},\ \bibinfo
  {year} {1971})\BibitemShut {NoStop}%
\end{thebibliography}
\end{document}